\documentclass[10pt]{amsart}

\newtheorem{theorem}{Theorem}[section]
\newtheorem{lemma}[theorem]{Lemma}           
\newtheorem{cor}[theorem]{Corollary}

\theoremstyle{definition}
\newtheorem{definition}[theorem]{Definition}

\theoremstyle{remark}
\newtheorem{remark}[theorem]{Remark}

\numberwithin{equation}{section}

\begin{document}

\author{Gregory Eskin}
\address{Department of mathematics, UCLA, Los Angeles, CA 90095-1555, USA}
\author{Hiroshi ISOZAKI}
\address{Institute of Mathematics \\
University of Tsukuba,
Tsukuba, 305-8571, Japan}
\author{Stephen O'Dell}
\address{Department of mathematics, UCLA, Los Angeles, CA 90095-1555, USA}

\title[Aharonov-Bohm Effect]{Gauge Equivalence and Inverse Scattering for Aharonov-Bohm Effect}


\maketitle

\begin{abstract}
We consider the Aharonov-Bohm effect for the Schr{\"o}dinger operator $H = \left(- i\nabla_x - A(x)\right)^2 + V(x)$ and the related inverse problem in an exterior domain $\Omega$ in ${\bf R}^2$ with Dirichlet boundary condition. We study the structure and asymptotics of generalized eigenfunctions and show that 
the scattering operator determines the domain $\Omega$ and $H$ up to gauge equivalence under the equal flux condition. We also show that the flux is determined by the scattering operator if the obstacle $\Omega^c$ is convex. 
\end{abstract}


\section{Introduction}


\subsection{Aharonov-Bohm Hamiltonian}
The aim of this paper is to study scattering phenomena of quantum mechanical particles governed by the Schr{\"o}dinger operator
\begin{equation}
H = \left(- i\nabla_x - A(x)\right)^2 + V(x)
\label{eq:Sec1ScrOpH}
\end{equation}
in an exterior domain $\Omega \subset {\bf R}^2$ with Dirichlet boundary condition. Our basic concern is the following situation. Given points $x^{(j)}, j = 1, \cdots, N$, we consider the magnetic field, which is identified with the 2-form
\begin{equation}
B(x)dx + \sum_{j=1}^N\alpha_j\delta(x - x^{(j)})dx, \quad \alpha_j \in {\bf R},
\nonumber
\end{equation}
where $dx = dx_1\wedge dx_2$. We put
\begin{equation}
\Theta(x,y) = \frac{(x-y)\times d\vec{x}}{|x-y|^2}, \quad
d\vec{x} = (dx_1,dx_2).
\nonumber
\end{equation}
Our motivating example for the magnetic vector potential is, identified with 1-form, 
\begin{equation}
A(x) = \frac{1}{2\pi}\int_{{\bf R}^2}\Theta(x,y)B(y)dy + \sum_{j=1}^N\alpha_j\Theta(x,x^{(j)}) + dL(x),
\label{eq:Sec1FormofA}
\end{equation}
where $|\partial_x^{\alpha}B(x)| \leq C_{\alpha}(1 + |x|)^{-2- |\alpha| -\epsilon_0}$  and 
$$
|\partial_x^{\alpha}L(x)| \leq C_{\alpha}(1 + |x|)^{-|\alpha| - \epsilon_0}, \quad\forall \alpha
$$
for some $\epsilon_0 > 0$. We take a small open set $\mathcal O_j$ containing $x^{(j)}$  so that $x^{(i)} \not\in \overline{\mathcal O_j}$ and $\mathcal O_i\cap\mathcal O_j = \emptyset$, if $i \neq j$. Let $\Omega = {\bf R}^2\setminus\cup_{j=1}^N\overline{\mathcal O_j}$. Note that the obstacle $\cup_{j=1}^N\overline{\mathcal O_j}$ is not convex if $N \geq 2$. When ${\rm supp}\,B(x) \subset \cup_{j=1}^N\mathcal O_j$, the magnetic fields are shielded. However, contrary to the intuition from classical electromagnetism, the particle {\it feels} the magnetic vector potential (\cite{AB59}, \cite{PeTo89}, \cite{Tetal86}).

This Aharonov-Bohm effect is a purely quantum mechanical phenomenon, affected by a topological nature of the domain $\Omega$. However, the long-range property of the associated magnetic vector potential makes it difficult to study the construction and  spatial asymptotics of distorted plane waves, and the mathematical works for the Aharonov-Bohm effect have been centered around 
the time-dependent scattering theory.

The spectrum and the singularities of the scattering operator were studied in \cite{RoYa02}, \cite{RoYa03} including the case of general long-range perturbations. In \cite{ItoTa01}, \cite{Tam06}, \cite{Tam07}, the asymptotics of the scattering matrix was computed in the semi-classical regime.


\subsection{Inverse problems}
In the works of Nicoleau \cite{Nic00}, and Weder \cite{We02}, the inverse problem was studied for the case of one convex obstacle : For two operators  $ (- i\nabla_x - A^{(j)}(x))^2$, $j = 1,2$, let $S(A^{(j)})$ be the scattering operator, and $\alpha^{(j)}$ the total flux of $A^{(j)}$. Then $S(A^{(1)}) = S(A^{(2)})$ implies $\alpha^{(1)} = \alpha^{(2)}$ mod 2, and 
$dA^{(1)} = dA^{(2)}$ on $\Omega$. We improve their result (cf. Theorem 5.9) by showing that $\alpha_1 = \alpha_2$ if $\alpha^{(i)}$ are not integers.

Alternatively, one can deal with the inverse problem for the wave equation 
$$
\left(\partial_t^2 + (-i\nabla_x - A(x))^2\right)u = 0
$$
 in a bounded domain using the hyperbolic Dirichlet-Neumann map (D-N map) instead of the scattering operator. In this case, one can apply the boundary control method (BC method) initiated by Belishev and developed by Belishev-Kurylev to identify the gauge equivalence classes of $A(x)$ and the domain $\Omega$ (see \cite{Be97}, \cite{BeKu92}, \cite{KuLa00}, \cite{KaKuLa01}). Some new ingredients of the BC method were also studied by \cite{Es06}, \cite{Es07}, where the emphasis was made on the gauge equivalence.

It is well-known that for short-range perturbations of $- \Delta$, the scattering matrix determines the D-N map in a bounded domain. Hence, the inverse scattering problem for the local perturabtion of $- \Delta$ can be reduced to the inverse boundary value problem, and one can apply the BC method to solve it. 
However, even if the magnetic field $B(x)$ has a compact support (this is the most interesting physical situation) the total Hamiltonian is a long-range perturbation of $- \Delta$. Indeed, suppose $B(x) = 0$ for $|x| > R$. Then there exists a magnetic potential $A(x)$ such that ${\rm curl}\,A(x) = B(x)$ and $A(x) = \alpha_0\Theta(x,x^{(0)}) + A'(x)$, where $A'(x) = 0$ for $|x| > R$, $x^{(0)} \in \Omega$ and $\alpha_0$ is the total flux. Assuming $\alpha_0 \neq 0$, we have that $A(x)$ is a long-range potential. Even in this case, it is not obvious how to determine the D-N map from the scattering operator. 
This is related to the fact that the asymptotic expansion of the distorted plane wave for the Aharonov-Bohm Hamiltonian was unknown.


\subsection{Main results}
In this paper, we shall deal with the case of $N \geq 1$ obstacles which are not necessarily convex. Our first main result is Theorem 5.7 which shows that $S(A^{(1)},V^{(1)}) = S(A^{(2)},V^{(2)})$ implies $\Omega^{(1)} = \Omega^{(2)} =:\Omega$, $V^{(1)} = V^{(2)}$ on $\Omega$ and 
$A^{(i)}$, $i = 1,2$, are gauge equivalent under the equal flux condition. The second main result is Theorem 5.9, which shows that if $\Omega^{(1)} = \Omega^{(2)}$, whose complement is convex, then the coincidence of the scattering operators implies that the fluxes are equal assuming that they are not integers.

Summarizing these two theorems, we get the following conclusion. 
To fix the idea, let us fix a domain $\Omega$ and a scalar electric potential $V(x)$. Then, Theorems 5.7 and 5.9 imply that, if $\Omega^c$ is convex, there is a 1 to 1 correspondence between equivalence classes of magnetic vector potentials and those of S-matrices, due to their gauge equivalences. This fact is also true for non-convex obstacles if we have the equal flux condition in Theorem 5.7. 
This equal flux condition is crucial. In fact, it is necessary for the existence of the above 1 to 1 correspondence (Theorem 5.10).

 We first use the results of \cite{Nic00}, \cite{We02}
 to show that if two scattering matrices coincide under equal flux condition, then the associated Schr{\"o}dinger operators are gauge equvalent near infinity.
 Note that this step is not needed when the magnetic field and electric potential have compact support. Next we use the spatial asymptotics of the distorted plane waves to derive the gauge equivalence of the D-N map for the boundary value problem. We emphasize that the known proofs for the case of short-range potential do not work here and one needs a more sophisticated technique developed in \cite{Ho85}, \cite{IsKi85b}, \cite{Is01} to get the result. From here we pass to the BC method to complete the proof of Theorem 5.7. The proof of Theorem 5.9 uses the estimates of singularties of the scattering matrix due to Roux and Yafaev \cite{RoYa02}, \cite{RoYa03}, \cite{Ya06}.


\subsection{Plan of the paper}
In \S 3 and \S 4, we study the stationary scattering theory for $H$.  In particular, Lemma 3.9 and Theorem 4.5 play key roles in the proof of Theorem 5.7. Another aim of \S 4 is to study the structure of distorted plane waves.
 When $B(x) = \alpha\delta(x)$, there exists an explicit solution $\psi_{AB}(x)$ to the Schr{\"o}dinger equation $\big((- i\nabla_x - A^{(0)}(x))^2 - \lambda\big)\psi_{AB} = 0$ proposed by Aharonov-Bohm (see also \cite{Rui83}).
We construct a distorted plane wave of $H$ containing $\psi_{AB}$ as its principal part, and study its asymptotic behavior at infinity in Lemmas 4.9, 4.10. They explain the relation between the scattering matrix and the phase of distorted plane waves. Although this result is not used directly in our procedure for  the inverse scattering, it is of independent interest since in the long-range scattering the construction and asymptotic expansion of distorted plane waves is no longer the same as the short-range case.

We use the following notation. For Banach spaces $X$ and $Y$, ${\bf B}(X;Y)$ denotes the totality of bounded operatos from $X$ to $Y$. For $a = (a_1,a_2), b = (b_1,b_2) \in {\bf C}^2$,
$$
a\times b = a_1b_2 - a_2b_1.
$$
 For $x \in {\bf R}^2$, we put 
$$
\langle x\rangle = (1 + |x|^2)^{1/2}, \quad 
\widehat x = x/|x|.
$$
 For a self-adjoint operator $H$, $\sigma_d(H)$, $\sigma_e(H)$ and $\sigma_p(H)$ denote the discrete spectrum, essential spectrum and point spectrum (= the set of all eigenvalues), respectively. $\mathcal H_{ac}(H)$ denotes the absolutely continuous subspace for $H$. For $f \in L^2({\bf R}^2)$, $\widehat f(\xi)$ denotes the Fourier transform of $f$: 
$$
\widehat f(\xi) = (2\pi)^{-1}\int_{{\bf R}^2}e^{-ix\cdot\xi}f(x)dx.
$$


\section{Resolvent estimates}


\subsection{Besov type spaces}
We define a Besov type space introduced  by Agmon-H{\"o}rmander \cite{AgHo76}. Let $\mathcal B$ be the Banach space of $L^2({\bf R}^2)$-functions equipped with norm 
\begin{equation}
\|f\|_{\mathcal B} = \sum_{j=0}^{\infty}2^{j/2}
\left(\int_{D_j}|f(x)|^2dx \right)^{1/2},
\nonumber
\end{equation}
where $D_0 = \{|x| < 1\}$, $D_j = \{2^{j-1} < |x| < 2^j\}, \ j \geq 1$. 
Its dual spce is identified with the set of $L^2_{loc}({\bf R}^2)$-functions $u(x)$ satisfying
\begin{equation}
\|u\|_{{\mathcal B}^{\ast}} = \sup_{R>1}\frac{1}{R}
\int_{|x|<R}|u(x)|^2dx < \infty.
\nonumber
\end{equation}
For $s \in {\bf R}$, the weighted $L^2$-space $L^{2,s}$ is defined by
\begin{equation}
u \in L^{2,s} \Longleftrightarrow \|u\|_s^2 = \int_{{\bf R}^2}(1 + |x|)^{2s}
|u(x)|^2dx < \infty.
\nonumber
\end{equation}
For $s > 1/2$, we have the following inclusion relations
\begin{equation}
L^{2,s} \subset \mathcal B \subset L^{2,1/2} \subset L^2 \subset L^{2,-1/2} \subset \mathcal B^{\ast} \subset L^{2,-s}.
\nonumber
\end{equation}
We use the notation $u \simeq v$, if
\begin{equation}
\lim_{R\to\infty}\frac{1}{R}\int_{|x|<R}|u(x) - v(x)|^2dx = 0.
\label{eq:Sec2EquivRel}
\end{equation}
The following lemma is easy to prove (see \cite{Is01}, Lemma 2.2).

\begin{lemma} 
For $u \in \mathcal B^{\ast}$, $u \simeq 0$ is equivalent to
$$
\lim_{R\to\infty}\frac{1}{R}\int \rho\big(\frac{|x|}{R}\big)|u(x)|^2dx = 0, \quad 
\forall \rho \in C_0^{\infty}((0,\infty)).
$$
\end{lemma}


\subsection{Resolvent estimates} Let $\Omega = {\bf R}^2\setminus\overline{\mathcal O}$ be a connected open set in ${\bf R}^2$ exterior to a bounded open set $\mathcal O$. We consider a Schr{\"o}dinger operator (\ref{eq:Sec1ScrOpH})
in $\Omega$ with Dirichlet boundary condition on $\partial\Omega$. The following assumptions are imposed on $H$. 

\medskip
\noindent
(A-1) The magnetic vector potential $A(x) = (A_1(x),A_2(x)) \in C^{\infty}(\overline{\Omega};{\bf R}^2)$ satisfies
\begin{equation}
|\partial_x^{\alpha}A(x)| \leq C_{\alpha}\langle x\rangle^{-1-|\alpha|}, \quad 
\forall \alpha,
\label{eq:Sect2Magnepot}
\end{equation}
and the transversal gauge condition
\begin{equation}
|\partial_x^{\alpha}\big(A(x)\cdot x\big)| \leq C_{\alpha}\langle x\rangle^{-1-|\alpha|}, \quad \forall \alpha.
\label{eq:Sect2transversal}
\end{equation}
(A-2) The magnetic field
\begin{equation}
B(x) = \frac{\partial A_2(x)}{\partial x_1} - \frac{\partial A_1(x)}{\partial x_2}
\nonumber
\end{equation} 
satisfies for some $\epsilon_0 > 0$
\begin{equation}
\left|\partial_x^{\alpha}\,B(x)\right| \leq C_{\alpha}\langle x\rangle^{- 2 - |\alpha| - \epsilon_0},  \quad \forall \alpha.
\label{eq:Sect2MagneEstimate}
\end{equation}
(A-3) The electric scalar potential $V(x) \in C^{\infty}(\overline{\Omega};{\bf R})$ satisfies
\begin{equation}
|\partial_x^{\alpha}V(x)| \leq C_{\alpha}\langle x\rangle^{- 1 - |\alpha|- \epsilon_0}.
\label{eq:Sect2Scalarpotential}
\end{equation}

\medskip
We summarize estimates of the resolvent $R(z) = (H - z)^{-1}$ in the following theorems. Note that the spaces $\mathcal B$ and $\mathcal B^{\ast}$ as well as $L^{2,s}$ are also defined on the domain $\Omega$. 


\begin{theorem} (1) $\ \sigma_d(H) \subset (-\infty,0)$, $\sigma_{e}(H) = [0,\infty)$. \\
\noindent
(2) $\ \sigma_p(H)\cap(0,\infty) = \emptyset$. \\
\noindent
(3) For any $\lambda > 0$ and $s > 1/2$, the following strong limit
\begin{equation}
\lim_{\epsilon\to 0}R(\lambda \pm i\epsilon)f =:R(\lambda \pm i0)f, \quad \forall f \in L^{2,s},
\nonumber
\end{equation}
exists in $L^{2,-s}$
and $(0,\infty) \ni \lambda \to R(\lambda \pm i0) \in {\bf B}(L^{2,s};L^{2,-s})$ is strongly continuous. For any $s > 1/2$ and compact interval $I \subset (0,\infty)$, there exists a constant $C_s > 0$ such that
\begin{equation}
\|R(\lambda \pm i0)f\|_{-s} \leq C_s\|f\|_{s}, \quad
\forall \lambda \in I.
\label{eq:ResolventL2s}
\end{equation}
(4) There exists $\alpha > - 1/2$ such that
\begin{equation}
\left(\nabla_x \mp i\sqrt{\lambda}\widehat x\right)R(\lambda \pm i0) \in {\bf B}(L^{2,s} ; L^{2,\alpha}),
\label{eq:RadCond1}
\end{equation}
\begin{equation}
\left(\nabla_x - \widehat x\frac{\partial}{\partial r}\right)R(\lambda \pm i0) \in {\bf B}(L^{2,s} ; L^{2,\alpha}).
\label{eq:RadCond2}
\end{equation}
\end{theorem}
Proof. The assertions (1), (2) are well-known. The assertion (3) and the estimate (\ref{eq:RadCond1}) are proved in \cite{IkSa72} for the whole space problem. It is not difficult to extend them to the exterior domain by a cutting-off argument. In fact, assuming that $\mathcal O \subset \{|x| < C_0\}$, we take $\chi(x) \in C^{\infty}({\bf R}^2)$  such that $\chi(x) = 0$ for $|x| < C_0 + 1$ and $\chi(x) = 1$ for $|x| > C_0 + 2$, and put $v = \chi(x)R(z)f$. Then $v$ satisfies
\begin{equation}
(H - z)v = \chi f + [H,\chi]R(z)f.
\label{eq:Sect2Cutoff}
\end{equation}
Therefore by (\ref{eq:ResolventL2s}) and the elliptic estimate, $u = R(z)f$ satisfies 
\begin{equation}
\|u\|_{-s} \leq C_s(\|f\|_s + \|u\|_{L^2(B)}),
\nonumber
\end{equation}
where $B$ is a bounded set in ${\bf R}^2$. Using this inequality, one can repeat the arguments in \cite{IkSa72} to obtain (3) and the estimate (\ref{eq:RadCond1}). Using 
$$
\nabla - \widehat x\frac{\partial}{\partial r} = \left(\nabla \mp i\sqrt{\lambda}\widehat x\right) - \widehat x\left(\frac{\partial}{\partial r} \mp i\sqrt{\lambda}\right),
$$
one can prove (\ref{eq:RadCond2}). \qed


\begin{theorem}
Suppose $u \in {\mathcal B}^{\ast}$ satisfies $(H - \lambda)u = 0$ $(\lambda > 0)$ in a neighborhood of infinity. Assume that
$$
\lim_{R\to\infty}\frac{1}{R}\int_{|x|<R}|u(x)|^2dx = 0.
$$
Then $u(x) = 0$ in a neighborhood of infinity.
\end{theorem}
Proof. By the assumption, we have
$$
\liminf_{r\to\infty}\, r\!\int_{S^1}|u(r\omega)|^2d\omega = 0.
$$
The theorem then follows from \cite{IkSa72}, Lemma 2.5. \qed


\begin{theorem}
For any compact interval $I \subset (0,\infty)$, there exists a constant $C > 0$ such that
$$
\|R(\lambda + i0)f\|_{{\mathcal B}^{\ast}} \leq C\|f\|_{\mathcal B}, \quad \lambda \in I.
$$
\end{theorem}
Proof. For ${\bf R}^2$, the proof is given in \cite{Ho85}, Theorem 30. 2. 10.
Alternatively, one can use Mourre's commutator method (\cite{JePe85}). To prove the theorem for $\Omega$, we first use (\ref{eq:Sect2Cutoff}) to see that 
$$
\|R(\lambda \pm i0)f\|_{{\mathcal B}^{\ast}} \leq C\|f\|_s, \quad s > 1/2.
$$
By taking the adjoint, we then have $R(\lambda \pm i0) \in {\bf B}(\mathcal B;L^{2,-s})$ $,s > 1/2$. Again using (\ref{eq:Sect2Cutoff}), we obtain 
$R(\lambda \pm i0) \in {\bf B}(\mathcal B;{\mathcal B}^{\ast})$.
\qed

\medskip
The following refinement of the radiation condition is also important.
A solution $u \in {\mathcal B}^{\ast}$ to the Schr{\"o}dinger equation $(H - \lambda)u = f, \ \lambda > 0$, is said to satisfy the outgoing radiation condition if it satisfies
\begin{equation}
\lim_{R\to\infty}\frac{1}{R}\int_{|x|<R}\big|\big(\frac{\partial}{\partial r} - i\sqrt{\lambda}\big)u(x)\big|^2dx = 0.
\label{eq:RadCond}
\end{equation}
If $i$ is replaced by $- i$, $u$ is said to satisfy the incoming radiation condition.


\begin{theorem}
(1) The solution $u \in \mathcal B^{\ast}$ of the equation $(H - \lambda)u = f \in {\mathcal B}$ satisfying the outgoing (or incoming) radiation condition is unique. \\
\noindent
(2) $ R(\lambda \pm i0)f$ is the unique solution of the equation $(H - \lambda)u = f \in \mathcal B$ satisfying the radiation condition (outgoing for $+$, incoming for $-$).
\end{theorem}
Proof. Suppose $u \in {\mathcal B}^{\ast}$ satisfies $(H - \lambda)u = 0$ and the outgoing radiation condition. Take a non-negative $\rho \in C_0^{\infty}((0,\infty))$ such that $\int_0^{\infty}\rho(t)dt = 1$, and put
$$
\varphi_R(x) = \chi\big(\frac{|x|}{R}\big), \quad
\chi(t) = \int_t^{\infty}\rho(s)ds.
$$
Since $\big((H-\lambda)u,\varphi_Ru\big) = 0$, by integrating by parts and taking the imaginary part, 
\begin{equation}
{\rm Re}\,\frac{1}{R}\big((i\partial_r + \widehat x\cdot A)u,\rho\big(\frac{r}{R}\big)u\big) = 0.
\nonumber
\end{equation}
We then have
\begin{equation}
\begin{split}
{\rm Re}\, \frac{1}{R}\left((-i\partial_r - \sqrt{\lambda})u,\rho\big(\frac{r}{R}\big)u\right)
+ \frac{\sqrt{\lambda}}{R}\left(u,\rho\big(\frac{r}{R}\big)u\right) 
= \frac{1}{R}\left(\widehat x\cdot Au,\rho\big(\frac{r}{R}\big)u\right).
\end{split}
\nonumber
\end{equation}
Let $R \to \infty$. Then the 1st term of the left-hand side vanishes by (\ref{eq:RadCond}), and so does the right-hand side by (\ref{eq:Sect2transversal}). Therefore $(u,\rho(r/R)u)/R \to 0$. Hence $u(x) = 0$ by Lemma 2.1 and Theorem 2.3, which proves (1).

To prove (2), we show that for $f \in {\mathcal B}$, $R(\lambda \pm i0)f$ satisfies the radiation condition. 
By Theorem 2.4, letting $u_{\pm} = R(\lambda \pm i0)f$, we have
\begin{equation}
\limsup_{R\to\infty}\frac{1}{R}\int_{|x|<R}\big|(\partial_r \mp i\sqrt{\lambda}\big)u_{\pm}\big|^2dx \leq
\|\big(\partial_r \mp i\sqrt{\lambda}\big)u_{\pm}\|^2_{{\mathcal B}^{\ast}} \leq C\|f\|^2_{\mathcal B}.
\nonumber
\end{equation}
If $f \in L^{2,s}, \ s > 1/2$, the left-hand side vanishes by virtue of (\ref{eq:RadCond1}). For $f \in {\mathcal B}$, we have only to approximate it by an element of $L^{2,s}$. \qed

\medskip
Let $\mathcal S^{\pm}$ be the set of symbols $p_{\pm}(x,\xi)$ satisfying
\begin{equation}
|\partial_x^{\alpha}\partial_{\xi}^{\beta}p_{\pm}(x,\xi)| \leq C_{\alpha\beta}\langle x\rangle^{-|\alpha|}\langle \xi\rangle^{-|\beta|}, \quad \forall \alpha, \beta,
\label{eq:S010symbol}
\end{equation}
and there exists a constant $- 1 < \mu_{\pm} < 1$, which is allowed to depend on $p_{\pm}(x,\xi)$, such that
\begin{equation}
p_{-}(x,\xi) = 0 \quad {\rm if} \quad \widehat x\cdot\widehat\xi > \mu_-, \quad p_{+}(x,\xi) = 0 \quad {\rm if} \quad \widehat x\cdot\widehat\xi < \mu_+.
\nonumber
\end{equation}
Since $\widehat x$ and $\widehat\xi$ should be well-defined, we are tacitly assuming that  $x$ and $\xi$ are non-zero on the support of the symbol of $p_{\pm}$.
For a pseudo-differential operator ($\Psi$DO) $P$, $P \in \mathcal S^{\pm}$ means that its symbol belongs to $\mathcal S^{\pm}$. The following theorem is proved in the same way as in \cite{IsKi85b}, Theorem 1, by using the parametrix in \cite{RoYa03}.


\begin{theorem}  (1) Let $\lambda > 0$ and $P$ be a $\Psi$DO such that its symbol $p(x,\xi)$ satisfies (\ref{eq:S010symbol}) and
$$
p(x,\xi) = 0, \quad {\rm if} \quad \frac{\lambda}{2} < |\xi|^2 < 2\lambda.
$$
Then 
$$
PR(\lambda \pm i0) \in {\bf B}(L^{2,s};L^{2,s}), \quad \forall s \geq 0.
$$
(2) Let $\lambda > 0$ and $P_{\pm} \in \mathcal S^{\pm}$. Then for any $s > 1/2$ and $\epsilon > 0$, we have
$$
P_{\mp}R(\lambda \pm i0) \in {\bf B}(L^{2,s};L^{2,s-1-\epsilon}).
$$
\end{theorem}


\begin{theorem}
Let $\lambda > 0$ and $P_{\pm}$ be such that its symbol satisfies (\ref{eq:S010symbol}) and
$$
p_-(x,\xi) = 0 \quad {\rm if} \quad \widehat x = \widehat\xi, \quad 
p_+(x,\xi) = 0 \quad {\rm if} \quad \widehat x = - \widehat\xi.
$$
 Let $s > 1/2$ be sufficiently close to $1/2$. Then there exists $\alpha > - 1/2$ such that
\begin{equation}
P_{\mp}R(\lambda \pm i0) \in {\bf B}(L^{2,s};L^{2,\alpha}).
\nonumber
\end{equation}
\end{theorem}
Proof. This theorem is essentially proved in \cite{Is01}, Theorem 3.5. For the reader's convenience, we reproduce the proof for the case $R(\lambda + i0)$.  Since $p_{-}(x,|\xi|\widehat x) = 0$, we have
\begin{equation}
\begin{split}
p_-(x,\xi) &= \int_0^1\frac{d}{dt}p_-(x,|\xi|(t\widehat\xi + (1 - t)\widehat x))dt \\
&= \int_0^1(\nabla_{\xi}p_-)(x,|\xi|(t\widehat \xi + (1 - t)\widehat x))dt\cdot(\widehat \xi - \widehat x).
\nonumber
\end{split}
\end{equation}
Therefore we have only to prove the theorem for the vector-valued symbol
$$
q(x,\xi) = \chi(x)\chi(\xi)(\widehat x - \widehat \xi),
$$
where $\chi \in C^{\infty}({\bf R}^2)$ such that $\chi(x) = 0$ for $|x| < \epsilon$, $\chi(x) = 1$ for $|x| > 2\epsilon$ for some $\epsilon > 0$. Take $\rho_{\pm}(t) \in C^{\infty}({\bf R})$ such that $\rho_+(t) + \rho_-(t) = 1$, $\rho_-(t) = 1 \ (t < - 1/2)$, $\rho_-(t) = 0 \ (t > 1/2)$ and split $q(x,\xi)$ into two parts :
$$
q(x,\xi) = \rho_+\big(\widehat x\cdot\widehat\xi\big)q(x,\xi) 
+ \rho_-\big(\widehat x\cdot\widehat\xi\big)q(x,\xi) =: q_+(x,\xi) + q_-(x,\xi).
$$
For the symbol $q_-$, the theorem is already proved. We put
$$
\nabla^{(s)} = \nabla_x - \widehat x\frac{\partial}{\partial r}.
$$
Taking notice of the relation
$$
|\xi - (\widehat x\cdot\xi)\widehat x|^2 = \frac{|\xi|^2}{2}(1 + \widehat x\cdot\widehat \xi\,)(\widehat x - \widehat \xi\,)^2,
$$
we have
$$
q_+(x,D_x)^{\ast}\langle x\rangle^{2\alpha}q_+(x,D_x) = (\nabla^{(s)})^{\ast}\langle x\rangle^{\alpha}P_0\langle x\rangle^{\alpha}\nabla^{(s)} + \langle x\rangle^{2\alpha-1}P_1,
$$
where $P_0$, $P_1$ are bounded $\Psi$DO's. Then the theorem readily follows from (\ref{eq:RadCond2}). \qed


\section{Spectral representation}


\subsection{Time-dependent scattering theory}
It is well-known that, although $H$ is a long-range perturbation of
$ - \Delta$, the usual wave operators exist (see \cite{LoTh87}). Since we need  a representation of the S-matrix by distorted plane waves (Lemma 3.9), we review relations between the usual wave operator and the modified wave operator.

We extend $A(x)$ smoothly on ${\bf R}^2$, and put
\begin{equation}
\Phi_{\pm}(x,\xi) = \mp \int_0^{\infty}A(x \pm s\xi)\cdot\xi \,ds,
\label{eq:Sect3Phixxi}
\end{equation}
and define
\begin{equation}
\varphi_{\pm}(x,\xi) = x\cdot\xi + \Phi_{\pm}(x,\xi).
\nonumber
\end{equation}
\begin{equation}
\left((-i\nabla_x - A(x))^2 - |\xi|^2\right)e^{i\varphi_{\pm}(x,\xi)} = e^{i\varphi_{\pm}(x,\xi)}q_{\pm}(x,\xi).
\nonumber
\end{equation}
For a small $0 < \delta < 1$, we define the region
\begin{equation}
D_{\delta}^{(\pm)} = \{(x,\xi) \in {\bf R}^2\times{\bf R}^2\, ;\, \pm\widehat x\cdot\widehat \xi \geq - 1 + \delta, \ |x| >\delta, \ |\xi| > \delta\}.
\nonumber
\end{equation}


\begin{lemma}
On $D_{\delta}^{(\pm)}$ we have the following estimates
\begin{equation}
|\partial_x^{\alpha}\partial_{\xi}^{\beta}\Phi_{\pm}(x,\xi)| \leq C_{\alpha\beta}\, \langle \xi\rangle^{-|\beta|}\langle x\rangle^{-|\alpha|}, \quad \forall \alpha, \beta.
\label{eq:Sect3Phiestimate}
\end{equation}
\begin{equation}
|\partial_x^{\alpha}\partial_{\xi}^{\beta}q_{\pm}(x,\xi)| \leq C_{\alpha\beta}\, \langle \xi\rangle^{-|\beta|}\langle x\rangle^{-2- \epsilon_0-|\alpha|}, \quad \forall \alpha, \beta.
\label{eq:Sect3qplusminus}
\end{equation}
 Moreover if $\widehat x = \pm \widehat \xi$, we have as $r = |x| \to \infty$,
\begin{equation}
\Phi_{\pm}(x,\xi) = O(r^{-1}).
\label{eq:Sect3Phivanish}
\end{equation}
\end{lemma}
Proof. Using the relation
\begin{equation}
\begin{split}
\left(A(x \pm s\xi) - A(\pm s\xi)\right)\cdot\xi = &
 \left(x_1\xi_2 - x_2\xi_1\right)\int_0^1B(\tau x \pm s\xi)d\tau \\
 & \pm \frac{d}{ds}\int_0^1x\cdot A(\tau x \pm s\xi)d\tau.
 \end{split}
\nonumber
\end{equation}
we have
\begin{equation}
\begin{split}
\Phi_{\pm}(x,\xi) = & \mp x\times\xi
 \int_0^{\infty}\int_0^1B(\tau x \pm s\xi)dsd\tau \\
 & + \int_0^1x\cdot A(\tau x)d\tau
 \mp\int_0^{\infty}A(\pm s\xi)\cdot\xi ds. 
 \end{split}
\label{eq:Sect3FormofPhi}
\end{equation}
The 2nd term of the right-hand side is rewritten as
\begin{equation}
\int_0^{\infty}\widehat x\cdot A(\rho \widehat x)d\rho - 
\int_r^{\infty}\rho\widehat x\cdot A(\rho \widehat x)\frac{d\rho}{\rho}.
\nonumber
\end{equation}
As $r \to \infty$, this behaves like a function of homogeneous degree 0 plus $O(r^{-1})$.
In the region $\{\pm \widehat x\cdot\widehat \xi \geq - 1 + \delta, \ |\xi| > \delta\}$
\begin{equation}
|\tau x \pm s\xi| \geq \frac{\sqrt\delta}{2}(\tau|x| + s|\xi|).
\nonumber
\end{equation}
Using this and (\ref{eq:Sect2MagneEstimate}), one can then prove (\ref{eq:Sect3Phiestimate}) by a direct computation. (\ref{eq:Sect3Phivanish}) is obvious. Differentiating (\ref{eq:Sect3Phixxi}), we have
\begin{equation}
\frac{\partial}{\partial x_1}\Phi_{\pm}(x,\xi) = 
\mp \xi_2\int_0^{\infty}B(x \pm s\xi)ds + A_1(x),
\label{eq:Sect3delPhidelx1}
\end{equation}
\begin{equation}
\frac{\partial}{\partial x_2}\Phi_{\pm}(x,\xi) = 
\pm \xi_1\int_0^{\infty}B(x \pm s\xi)ds + A_2(x).
\label{eq:Sect3depPhidelx2}
\end{equation}
By a direct computation, we have
\begin{equation}
q_{\pm}(x,\xi) = |\nabla_x\Phi_{\pm} - A|^2  - i\nabla_x\cdot(\nabla_x\Phi_{\pm} - A),
\nonumber
\end{equation}
where we have used the fact that $\xi\cdot(\nabla_x\Phi_{\pm} - A) = 0$ by (\ref{eq:Sect3delPhidelx1}) and (\ref{eq:Sect3depPhidelx2}).
Using (\ref{eq:Sect2MagneEstimate}) and (\ref{eq:Sect3delPhidelx1}), (\ref{eq:Sect3depPhidelx2}), we obtain (\ref{eq:Sect3qplusminus}).
\qed

\bigskip
We put
\begin{equation}
J_{\pm}f(x) = (2\pi)^{-n/2}\int_{{\bf R}^{n}}e^{i\varphi_{\pm}(x,\xi)}\chi_{\pm}(\widehat x\cdot\widehat\xi)\chi_{\Omega}(x)\widehat f(\xi)d\xi,
\label{eq:Jplusminus}
\end{equation}
where $\chi_{\pm}(t) \in C^{\infty}({\bf R})$ such that $\chi_+(t) = 1$ for $t > - 1 + 2\delta$,  $\chi_+(t) = 0$ for $t < - 1 + \delta$, and $\chi_-(t) = \chi_{+}(-t)$, and $\chi_{\Omega}(x) \in C^{\infty}({\bf R}^{2})$ such that $\chi_{\Omega}(x) = 1$ for $|x| > 2R$, $\chi_{\Omega}(x) = 0$ for $|x| < R$, $R$ being a constant satisfying $\overline{\mathcal O} \subset \{|x| < R\}$.
Define the modified wave operator $\mathcal M_{\pm}$ by
\begin{equation}
\mathcal M_{\pm} = \mathop{\rm s-lim}_{t\to\pm\infty}
e^{itH}J_{\pm}e^{-itH_0},
\label{eq:Sec3ModifiWaveOp1}
\end{equation}
where $H_0 = 
- \Delta_x$ in ${\bf R}^2$.


\begin{theorem}
The strong limit (\ref{eq:Sec3ModifiWaveOp1}) exists on $L^2({\bf R}^2)$, and is unitary from $L^2({\bf R}^2)$ onto ${\mathcal H}_{ac}(H)$. It has the intertwining property: $\varphi(H)\mathcal M_{\pm} = \mathcal M_{\pm}\varphi(H_0)$, where $\varphi$ is any bounded Borel function on ${\bf R}$. 
\end{theorem}

This theorem is proved in the same way as \cite{IsKi85a}, Theorem 1.1, \cite{Nic00}, Theorem 4 or \cite{RoYa03}, Theorem 5.10. 


\begin{theorem}
The usual wave operator
\begin{equation}
W_{\pm} = \mathop{\rm s-lim}_{t\to\pm\infty}
e^{itH}r_{\Omega}e^{-itH_0},
\label{eq:Sec3WaveOp2}
\end{equation}
exists and is equal to the modified wave operator $\mathcal M_{\pm}$, where $r_{\Omega}$ is the operator of restriction to $\Omega$, 
\end{theorem}
Proof. Using the stationary phase method and (\ref{eq:Sect3Phivanish}), we have for any $\widehat f(\xi) \in C_0^{\infty}({\bf R}^2\setminus\{0\})$,
$$
J_{\pm}e^{-itH_0}f \sim C|t|^{-1}e^{i|x|^2/(4t)}
\widehat f\big(\frac{x}{2t}\big) \sim e^{-itH_0}f
$$
as $t \to \pm \infty$, which together with Theorem 3.2 proves the theorem. \qed


\subsection{Spectral representation}
For $\lambda > 0$, we put
\begin{equation}
\left(\widetilde{\mathcal F_{0\pm}}(\lambda)f\right)(\omega) = 
\frac{1}{2\sqrt2\,\pi}\int_{{\bf R}^2}
e^{-i\varphi_{\pm}(x,\sqrt{\lambda}\omega)}
\chi_{\pm}(\widehat x\cdot\omega)\chi_{\Omega}(x)f(x)dx,
\label{eq:SectmathcalF0lmabda)}
\end{equation}
with $\chi_{\pm}$ and $\chi_{\Omega}$ as above.


\begin{lemma}
For any $\delta > 0$, there exists a constant $C = C_{\delta} > 0$ such that
\begin{equation}
\|\widetilde{\mathcal F_{0\pm}}(\lambda)f\|_{L^2(S^{1})} \leq C\|f\|_{\mathcal B}, \quad 
\forall \lambda > \delta.
\nonumber
\end{equation}
\end{lemma}
Proof. We put $a_{\pm}(x,\xi) = 
e^{-i\Phi_{\pm}(x,\xi)}\chi_{\pm}(\widehat x\cdot\widehat \xi)\chi_{\Omega}(x)
\chi_0(\xi)$, where $\chi_0(\xi) \in C_0^{\infty}({\bf R}^2)$ such that $\chi_0(\xi) = 1$ if $|\xi|^2 > \delta$ and $\chi_0(\xi) = 0$ if $|\xi|^2 < \delta/2$. Let $A_{\pm}$ be the $\Psi$DO defined by
$$
A_{\pm}f(x) = 2^{-1/2}(2\pi)^{-2}\iint e^{i(x-y)\cdot\xi}a_{\pm}(y,\xi)f(y)dyd\xi.
$$
Then $\widetilde{\mathcal F_{0\pm}}(\lambda)f = (\widehat{A_{\pm}f})(\sqrt\lambda\omega)$. 
Since $A \in {\bf B}(\mathcal B;\mathcal B)$ by \cite{AgHo76}, Theorem 2.5, the lemma follows. \qed

\bigskip
We put
\begin{equation}
(H - |\xi|^2)e^{i\varphi_{\pm}(x,\xi)}\chi_{\pm}(\widehat x\cdot\widehat\xi)\chi_{\Omega}(x) = e^{i\varphi_{\pm}(x,\xi)}g_{\pm}(x,\xi),
\label{eq:Sect3gplusminus}
\end{equation}
and define an operator $\mathcal G_{\pm}(\lambda)$ by
\begin{equation}
\left(\mathcal G_{\pm}(\lambda)f\right)(\omega) = \frac{1}{2\sqrt2\pi}\int_{\bf R^2}
e^{-i\varphi_{\pm}(x,\sqrt{\lambda}\omega)}
\overline{g_{\pm}(x,\sqrt{\lambda}\omega)}f(x)dx.
\label{eq:Sect3Glambda}
\end{equation}
We finally define
\begin{equation}
\mathcal F^{(\pm)}(\lambda) = \widetilde{\mathcal F_{0\pm}}(\lambda) - \mathcal G_{\pm}(\lambda)R(\lambda \pm i0).
\label{eq:Sect3MathcalFplusminus}
\end{equation}

\begin{lemma}
Let $I$ be any compact interval in $(0,\infty)$. Then for $s > 1/2$, there exists a constant $C > 0$ such that
\begin{equation}
\|\mathcal F^{(\pm)}(\lambda)f\|_{L^2(S^{1})} \leq C\|f\|_{s}, \quad 
\forall \lambda \in I.
\nonumber
\end{equation}
\end{lemma}
Proof. We consider the case of $\mathcal F^{(+)}(\lambda)$. By Lemma 3.4, $\widetilde{\mathcal F_{0+}}(\lambda)$ has the desired property. Let $\chi(\lambda) \in C_0^{\infty}({\bf R})$ be such that $\chi(\lambda) = 1$ on $I$ and $\chi(\lambda) = 0$ outside a small neighborhood of $I$. By Theorem 2.6 (1), one can insert $\chi(H_0)$ between $\mathcal G_{+}(\lambda)$ and $R(\lambda + i0)$. We next decompose the phase space according to the value of $\widehat x\cdot\widehat\xi$. More precisely, we consider $\Psi$DO's $P_{+}$ and $P_-$ with symbol $\widetilde\chi_+(\widehat x\cdot\widehat\xi)\chi(|\xi|^2)$ and $(1 - \widetilde\chi_+(\widehat x\cdot\widehat\xi))\chi(|\xi|^2)$, respecively, where $\widetilde\chi(t) = 1$ for $t > -1 + 3\delta$, $\widetilde\chi(t) = 0$ for $t < -1+2\delta$.
 By Lemma 3.1, $g_{+}(x,\xi) = O(|x|^{-1-\epsilon_0})$ on the support of the symbol of $P_{+}$. Therefore $\mathcal G_{+}(\lambda)P_{+}R(\lambda + i0) \in {\bf B}(L^{2,s};L^2(S^{1}))$. Since $g_{+}(x,\xi)$ contains $\nabla_x\chi_{+}(\widehat x\cdot\widehat\xi)$, $g_{+}(x,\xi) = O(|x|^{-1})$ on the support of the symbol of $P_{-}$. 
However, by Theorem 2.6 (2) we see that $P_{-}R(\lambda + i0) \in {\bf B}(L^{2,s};L^{2,\alpha})$ for some $\alpha > - 1/2$. 
Therefore $\mathcal G_{+}(\lambda)P_{-}R(\lambda + i0) \in {\bf B}(L^{2,s};L^2(S^1))$.
\qed


\begin{theorem} (1) The operator $\left(\mathcal F^{(\pm)}f\right)(\lambda,\omega) = \left(\mathcal F^{(\pm)}(\lambda)f\right)(\omega)$, defined for $f \in L^{2,s} \ (s > 1/2)$, is uniquely extended to a partial isometry with intial set ${\mathcal H}_{ac}(H)$ and final set $L^2((0,\infty);L^2(S^{1}); d\lambda)$. \\
\noindent
(2) For $f \in D(H)$, $\left(\mathcal F^{(\pm)}Hf\right)(\lambda) = \lambda \left(\mathcal F^{(\pm)}f\right)(\lambda)$. \\
\noindent
(3) $\mathcal F^{(\pm)}(\lambda)^{\ast} \in {\bf B}(L^2(S^{1});\mathcal B^{\ast})$ is an eigenoperator of $H$ in the sense that
$$
(H - \lambda)\mathcal F^{(\pm)}(\lambda)^{\ast}\phi = 0, \quad \forall \phi \in L^2(S^{1}).
$$
(4) For any $0 < a < b < \infty$ and $g \in L^2((0,\infty);L^2(S^{1}); d\lambda)$,
$$
\int_a^b\mathcal F^{(\pm)}(\lambda)^{\ast}g(\lambda)\, d\lambda \in
L^2({\Omega}).
$$
Moreover for any $f \in \mathcal H_{ac}(H)$, the following inversion formula holds :$$
f = \mathop{\rm s-lim}_{a\to0,b\to\infty}\int_a^b\mathcal F^{(\pm)}(\lambda)^{\ast}\left(\mathcal F^{(\pm)}f\right)(\lambda)\,  d\lambda.
$$
\end{theorem}

Proof. Since this theorem is well-known, we only give the sketch of the proof. Let $J_{\pm}$ be as in (\ref{eq:Jplusminus}).
We also put
\begin{equation}
G_{\pm}f(x) = (2\pi)^{-1}\int_{{\bf R}^2}e^{i\varphi_{\pm}(x,\xi)}
g_{\pm}(x,\xi)\widehat f(\xi)\,d\xi,
\nonumber
\end{equation}
where $g_{\pm}$ is defined by (\ref{eq:Sect3gplusminus}). 
Using the relation
\begin{equation}
HJ_{\pm} - J_{\pm}H_0 = G_{\pm},
\nonumber
\end{equation}
we have for $\widehat f \in C_0^{\infty}({\bf R}^2\setminus\{0\})$ and $g \in C_0^{\infty}(\Omega)$,
\begin{equation}
\begin{split}
(\mathcal M_{\pm}f,g) &= (J_{\pm}f,g) + i\int_0^{\pm\infty}(e^{itH}G_{\pm}e^{-itH_0}f,g)\,dt \\
&= (f,J_{\pm}^{\ast}g) - \int_{-\infty}^{\infty}(f,E'_0(\lambda)G_{\pm}^{\ast}R(\lambda \pm i0)g)\,d\lambda,
\end{split}
\nonumber
\end{equation}
where 
$$
E'_0(\lambda) = \frac{1}{2\pi i}\left(R_0(\lambda + i0) - R_0(\lambda - i0)\right).
$$
Letting
\begin{equation}
\left(\mathcal F_0(\lambda)f\right)(\omega) = (2\sqrt2\,\pi)^{-1}\int_{{\bf R}^2}e^{-i{\sqrt\lambda}\omega\cdot x}f(x)dx,
\label{eq:FourierF0}
\end{equation}
we then have
\begin{equation}
(\mathcal M_{\pm}f,g) = (f,J_{\pm}^{\ast}g) - \int_{0}^{\infty}(\mathcal F_0(\lambda)f,\mathcal F_0(\lambda)G_{\pm}^{\ast}R(\lambda \pm i0)g)\,d\lambda.
\nonumber
\end{equation}
The operator $\big(\mathcal F_0f\big)(\lambda,\omega) = \big(\mathcal F_0(\lambda)f\big)(\omega)$ is uniquely extended to a unitary from $L^2({\bf R}^2)$ to $L^2((0,\infty);L^2(S^{1});d\lambda)$. Therefore, in view of (\ref{eq:SectmathcalF0lmabda)}) and (\ref{eq:Sect3MathcalFplusminus}),  we have $\mathcal F^{(\pm)} = \mathcal F_0\mathcal M_{\pm}^{\ast}$.
By Therorem 3.3, this is equal to $\mathcal F_0W_{\pm}^{\ast}$. We have thus proven


\begin{lemma}
$\ 
\mathcal F^{(\pm)} = \mathcal F_0\left(W_{\pm}\right)^{\ast}.
$
\end{lemma}

By this lemma, $\mathcal F^{(\pm)}$ is a partial isometry with initial set $\mathcal H_{ac}(H)$ and final set $L^2((0,\infty);L^2(S^{1});d\lambda)$. The intertwining property of the wave operator implies Theorem 3.6 (2), and also for any compact interval $I \subset (0,\infty)$
\begin{equation}
\int_I(\mathcal F^{(\pm)}(\lambda)f,\mathcal F^{(\pm)}(\lambda)g)d\lambda = \frac{1}{2\pi i}\int_I((R(\lambda + i0) - R(\lambda -i0))f,g)d\lambda.
\nonumber
\end{equation}
Differentiating this, we have
\begin{equation}
(\mathcal F^{(\pm)}(\lambda)f,\mathcal F^{(\pm)}(\lambda)g) = \frac{1}{2\pi i}((R(\lambda + i0) - R(\lambda -i0))f,g).
\label{eq:Parseval}
\end{equation}
This and Theorem 2.1 imply


\begin{lemma}
$$
\mathcal F^{(\pm)}(\lambda) \in {\bf B}(\mathcal B;L^2(S^{1})).
$$
\end{lemma}
 The proof of the other assertions of Theorem 3.6 is standard and is omitted. \qed


\subsection{S-matrix}
The scattering operator $S$ is defined by
\begin{equation}
S = \big(W_+\big)^{\ast}W_-.
\nonumber
\end{equation}
By Lemma 3.7, its Fourier transform $\widehat S := \mathcal F_0S\big(\mathcal F_0\big)^{\ast}$ is written as
\begin{equation}
\widehat S = \mathcal F^{(+)}\big(\mathcal F^{(-)}\big)^{\ast}.
\nonumber
\end{equation}
As is well-known, it admits a diagonal representation:
\begin{equation}
\big(\widehat Sf\big)(\lambda,\omega) = \big(\widehat S(\lambda)f(\lambda,\cdot)\big)(\omega), \quad \forall f \in L^2((0,\infty);L^2(S^1);d\lambda),
\quad \omega \in S^1,
\nonumber
\end{equation}
where $\widehat S(\lambda)$ is a unitary operator on $L^2(S^1)$, called the S-matrix. 
It has the following expression.


\begin{lemma}
For $\lambda > 0$ and $\phi \in C^{\infty}(S^1)$, we have
\begin{equation}
\widehat S(\lambda)\phi = - 2\pi i\mathcal F^{(+)}(\lambda)\mathcal G_-(\lambda)^{\ast}\phi.
\nonumber
\end{equation}
\end{lemma}
Proof. First we make a comment on the above expression of $\widehat S(\lambda)$. By (\ref{eq:Sect3gplusminus}), $g_-(x,\xi)$ contains a factor $\nabla_x\chi_-(\widehat x\cdot\widehat\xi)$, which is $O(|x|^{-1})$. However, the stationary phase method implies
$$
\int_{S^{1}}e^{i\varphi_-(x,\sqrt{\lambda}\omega)}\nabla_x\chi_-(\widehat x\cdot\omega)\phi(\omega)d\omega = O(|x|^{-{\infty}}).
$$
We then have $\mathcal G_-(\lambda)^{\ast}\phi \in L^{2,s}$ with $s > 1/2$. Hence $\mathcal F^{(+)}(\lambda)\mathcal G_-(\lambda)^{\ast} \phi$ is well-defined.

We now prove the lemma. We put
\begin{equation}
J = J_+ + J_-.
\nonumber
\end{equation}
Then since $J_{\mp}e^{-itH_0} \to 0$ as $t \to \pm \infty$, we have
\begin{equation}
W_{\pm} = \mathop{\rm s-lim}_{t\to\pm \infty}e^{itH}Je^{-itH_0}.
\nonumber
\end{equation}
Letting
$$
G = G_+ + G_-,
$$
we have $HJ - JH_0 = G$, hence
$$
W_{\pm} = J + i\int_0^{\pm\infty}e^{itH}Ge^{-itH_0}ds.
$$
This yields
$$
W_+ - W_- = i\int_{-\infty}^{\infty}e^{itH}Ge^{-itH_0}ds.
$$
Since $S - 1 = \left(W_+\right)^{\ast}\left(W_- - W_+\right)$, we have
\begin{equation}
\begin{split}
(Sf,g) - (f,g) =& - i\int_{-\infty}^{\infty}\left(e^{itH}Ge^{-itH_0}f,W_+g\right)dt \\
 =& - i\int_{-\infty}^{\infty}\left(Ge^{-itH_0}f,W_+e^{-itH_0}g\right)dt \\
 =& - i\int_{-\infty}^{\infty}\left(Ge^{-itH_0}f,J_+e^{-itH_0}g\right)dt \\
& - \int_0^{\infty}ds\int_{-\infty}^{\infty}
\left(Ge^{-itH_0}f,e^{isH}G_+e^{-i(s+t)H_0}g\right)dt,
\end{split}
\label{eq:Sect3S-1}
\end{equation}
where we have used $e^{-itH}W_+ = W_+ e^{-itH_0}$ in the 2nd line and
$$
W_+ = J_+ + i\int_0^{\infty}e^{isH}G_+e^{-isH_0}ds
$$
in the 3rd line. Letting 
$\widehat f(\lambda) = \mathcal F_0(\lambda)f$, $\widehat g(\lambda) = \mathcal F_0(\lambda)g$, we have
\begin{equation}
\begin{split}
\int_{-\infty}^{\infty}
 & \left(G_+^{\ast}e^{-isH}Ge^{-itH_0}f,e^{-i(s+t)H_0}g\right)dt \\
=& \int_{-\infty}^{\infty}dt\int_0^{\infty}
\left(\mathcal F_0(\lambda)G_+^{\ast}e^{-isH}Ge^{-itH_0}f,e^{-i(s+t)\lambda}
\widehat g(\lambda)\right)d\lambda.
\end{split}
\nonumber
\end{equation}
Inserting $e^{-\epsilon|t|}$, and letting $\epsilon \to 0$, this converges to
\begin{equation}
\begin{split}
 & 2\pi\int_0^{\infty}\left(\mathcal F_0(\lambda)G_+^{\ast}e^{-is(H-\lambda)}
 GE_0'(\lambda)f,\widehat g(\lambda)\right)d\lambda \\
 & =  \, 2\pi\int_0^{\infty}\left(\mathcal F_0(\lambda)G_+^{\ast}e^{-is(H-\lambda)} G\mathcal F_0(\lambda)^{\ast}\widehat f(\lambda),\widehat g(\lambda)\right)d\lambda,
\end{split}
\nonumber
\end{equation}
where $E_0'(\lambda) = \frac{1}{2\pi i}\left(R_0(\lambda + i0) - R_0(\lambda - i0)\right) = \mathcal F_0(\lambda)^{\ast}\mathcal F_0(\lambda)$. Therefore the last term of the right-han side of (\ref{eq:Sect3S-1}) is equal to
$$
- 2\pi\int_0^{\infty}ds\int_0^{\infty}\left(\mathcal F_0(\lambda)G_+^{\ast}e^{-is(H-\lambda)} G\mathcal F_0(\lambda)^{\ast}\widehat f(\lambda),\widehat g(\lambda)\right)d\lambda.
$$
Inserting $e^{-\epsilon s}$ and letting $\epsilon \to 0$, this converges to
$$
2\pi i\int_0^{\infty}\left(\mathcal F_0(\lambda)G_+^{\ast}R(\lambda + i0)G\mathcal F_0(\lambda)^{\ast}\widehat f(\lambda),\widehat g(\lambda)\right)d\lambda.
$$
Similarly, the 1st term of the right-hand side of (\ref{eq:Sect3S-1}) is rewritten as
$$
- 2\pi i\int_0^{\infty}\left(\mathcal F_0(\lambda)J_+^{\ast}G\mathcal F_0(\lambda)^{\ast}\widehat f(\lambda),\widehat g(\lambda)\right)d\lambda.
$$
The above computations are justified when $\widehat f(\lambda), \widehat g(\lambda) \in C_0^{\infty}((0,\infty);L^2(S^1))$. 
We have thus proven that
$$
\widehat S(\lambda) = 1 - 2\pi i
\mathcal F_0(\lambda)\left(J_+^{\ast}G - G_+^{\ast}R(\lambda + i0)G\right)\mathcal F_0(\lambda)^{\ast}.
$$
By (\ref{eq:SectmathcalF0lmabda)}) and (\ref{eq:Sect3Glambda}), we have 
\begin{equation}
\widetilde{\mathcal F_{0\pm}}(\lambda) = \mathcal F_0(\lambda)J_{\pm}^{\ast},
\quad
\mathcal G_{\pm}(\lambda) = \mathcal F_0(\lambda)G_{\pm}^{\ast}.
\nonumber
\end{equation}
This implies
\begin{equation}
\begin{split}
\widehat S(\lambda) =  1 &- 2\pi i\, \widetilde{\mathcal F_{0+}}(\lambda)\Big(\mathcal G_+(\lambda)^{\ast} + \mathcal G_-(\lambda)^{\ast}\Big) \\
& + 2\pi i\, \mathcal G_+(\lambda)R(\lambda + i0)
\Big(\mathcal G_+(\lambda)^{\ast} + \mathcal G_-(\lambda)^{\ast}\Big).
\end{split}
\end{equation}
Here let us note that for $\phi \in C^{\infty}(S^1)$, $\widetilde{\mathcal F_{0+}}(\lambda)^{\ast}\phi$ satisfies the outgoing radiation condition, and
\begin{equation}
(H - \lambda)\widetilde{\mathcal F_{0+}}(\lambda)^{\ast}\phi = 
\mathcal G_+(\lambda)^{\ast}\phi.
\label{eq:S3HmathcalF0lambdaast}
\end{equation}
Theorem 2.5 then implies
$$
R(\lambda + i0)\mathcal G_+(\lambda)^{\ast}\phi = 
\widetilde{\mathcal F_{0+}}(\lambda)^{\ast}\phi.
$$
In view of (\ref{eq:Sect3MathcalFplusminus}), we have
\begin{equation}
\begin{split}
\widehat S(\lambda) =& 1 - 2\pi i\left(\widetilde{\mathcal F_{0+}}(\lambda)
\mathcal G_+(\lambda)^{\ast} - \mathcal G_+(\lambda)\widetilde{\mathcal F_{0+}}(\lambda)^{\ast}\right)  - 2\pi i \mathcal F^{(+)}(\lambda)\mathcal G_-(\lambda)^{\ast}.
\end{split}
\nonumber
\end{equation}
The proof of the lemma will then be completed if we show
\begin{equation}
2\pi i\left(\widetilde{\mathcal F_{0+}}(\lambda)
\mathcal G_+(\lambda)^{\ast} - \mathcal G_+(\lambda)\widetilde{\mathcal F_{0+}}(\lambda)^{\ast}\right) = 1.
\label{eq:Sec3FGrelation}
\end{equation}
For $\phi, \psi \in C^{\infty}(S^1)$, we put $u = \widetilde{\mathcal F_{0+}}(\lambda)^{\ast}\phi$, $v = \widetilde{\mathcal F_{0+}}(\lambda)^{\ast}\psi$. Then by the stationary phase method, we have as $r = |x| \to \infty$
\begin{equation}
u \sim \frac{e^{-\pi i/4}}{2\sqrt{\pi}\lambda^{1/4}}r^{-1/2}e^{i\sqrt{\lambda}r}\phi(\widehat x). 
\nonumber
\end{equation}
Then we have by integration by parts
\begin{equation}
\begin{split}
 \lim_{r\to\infty}\int_{|x|<r}\left((H-\lambda)u\overline{v} - u\overline{(H - \lambda)v}\right)dx 
= & \lim_{r\to\infty} - \int_{|x|=r}
\left(\frac{\partial u}{\partial r}\overline{v} - 
u\frac{\partial \overline{v}}{\partial r}\right)dS \\
= & -2i\sqrt{\lambda}\lim_{r\to\infty}\int_{|x|=r}u\overline{v}dS \\
= & \frac{1}{2\pi i}(\phi,\psi)_{L^2(S^1)},
\end{split}
\nonumber
\end{equation}
which proves (\ref{eq:Sec3FGrelation}) by (\ref{eq:S3HmathcalF0lambdaast}). \qed


\section{Distorted plane waves}
The main result of this section is Theorem 4.4 on the asymptotic expansion of the resolvent at infinity. With the aid of this theorem, we shall derive the asymptotic expansion of distorted plane waves (Theorem 4.5 and Lemma 4.10).


\subsection{Asymptotic expansion of the resolvent}


\begin{lemma}
Let $\rho(t) \in C_0^{\infty}((0,\infty))$ be such that $\int_0^{\infty}\rho(t)dt = 1$. Then for any $f \in L^{2,s}$ with $s > 1/2$ we have 
\begin{equation}
\lim_{R\to\infty}\mp \frac{i}{\pi R}\sqrt{\frac{\lambda}{2}}\int_{{\bf R}^2}
e^{-i\sqrt{\lambda}\omega\cdot x}\rho\big(\frac{|x|}{R}\big)R(\lambda \pm i0)f\,dx = \mathcal F^{(\pm)}(\lambda)f
\nonumber
\end{equation}
in the sense of strong limit  in $L^2(S^1)$.
\end{lemma}
Proof. We first consider $e^{-i\varphi_{\pm}(x,\xi)}$ instead of $e^{-ix\cdot\xi}$. Let $\rho_1(t) = \int_t^{\infty}\rho(s)ds$, and put $u_{\pm} = R(\lambda \pm i0)f$. Then we have
\begin{equation}
\begin{split}
& \int \left[(H -\lambda)e^{-i\varphi_{\pm}(x,\sqrt\lambda\omega)}\chi_{\pm}(\widehat x\cdot\omega)\chi_{\Omega}(x)\right]\rho_1\big(\frac{r}{R}\big)u_{\pm}dx \\
& = \int e^{-i\varphi_{\pm}(x,\sqrt\lambda\omega)}\overline{g_{\pm}(x,\sqrt{\lambda}\omega)}\rho_1\big(\frac{r}{R}\big)u_{\pm}dx,
\end{split}
\label{eq:Sect4Greenformula}
\end{equation}
where $r = |x|$. By integration by parts, the left-hand side is equal to
\begin{equation}
\int e^{-i\varphi_{\pm}(x,\sqrt\lambda\omega)}\chi_{\pm}(\widehat x\cdot\omega)\chi_{\Omega}(x)\big(H - \lambda\big)\rho_1\big(\frac{r}{R}\big)u_{\pm}dx 
\nonumber
\end{equation}
We compute
\begin{equation}
\begin{split}
\big(H - \lambda)\rho_1u_{\pm}  = & \rho_1f \pm \frac{2i\sqrt{\lambda}}{R}\rho\big(\frac{r}{R}\big)u_{\pm} \\
& + \frac{2}{R}\rho\big(\frac{r}{R}\big)\left(\frac{\partial}{\partial r} \mp i\sqrt{\lambda}\right)u_{\pm} - (\Delta\rho_1)u_{\pm}  + 2i(A\cdot\nabla\rho_1)u_{\pm}.
\end{split}
\nonumber
\end{equation}
By Theorem 2.2, the last 3 terms of the right-hand side tends to 0 in $L^{2,s'}$ for some $s' > 1/2$.
We have, therefore, letting $R \to \infty$ in (\ref{eq:Sect4Greenformula}),
\begin{equation}
\lim_{R\to\infty}\mp \frac{i}{\pi R}\sqrt{\frac{\lambda}{2}}\int_{{\bf R}^2}
e^{-i\varphi_{\pm}(x,\sqrt{\lambda}\omega)}\rho\big(\frac{|x|}{R}\big)
\chi_{\pm}(\widehat x\cdot \omega)R(\lambda \pm i0)f\,dx = \mathcal F^{(\pm)}(\lambda)f.
\label{eq:Sect4limitresoltoF}
\end{equation}
 Take $\chi_0(\xi) \in C_0^{\infty}({\bf R}^2)$ such that $\chi_0(\xi) = 0$ for $|\xi| < \sqrt{\lambda}/2$ or $|\xi| > 2\sqrt{\lambda}$ and $\chi_0(\xi) = 1$ near $|\xi| = \sqrt{\lambda}$.
Let $v^{(\pm)}_R = R^{-1}\rho(r/R)u_{\pm}$. Then $(1 - \chi_0(D_x))v_R^{(\pm)} \to 0$ in $\mathcal B$ by Lemma 2.2. Let $t_{\pm}(x,\omega) = 1 - e^{-i\Phi_{\pm}(x,\sqrt{\lambda}\omega)}\chi_{\pm}(\widehat x\cdot\omega)$ and consider the $\Psi$DO $T_{\pm}$ such that
$$
T_{\pm}f(x) = (2\pi)^{-1}\iint e^{i(x-y)\cdot\xi}t_{\pm}(y,\widehat\xi\,)\chi_0(\xi)f(y)dyd\xi.
$$
Since
\begin{equation}
\begin{split}
t_{\pm}(x,\omega) = & 1 - e^{-i\Phi_{\pm}(x,\pm \sqrt{\lambda}\widehat x)} - 
\left(e^{-i\Phi_{\pm}(x,\sqrt{\lambda}\omega)} - e^{-i\Phi_{\pm}(x,\pm\sqrt{\lambda}\widehat x)}\right)\chi_{\pm}(\widehat x\cdot\omega) \\
 & - e^{-i\Phi_{\pm}(x,\pm\sqrt{\lambda}\widehat x)}(\chi_{\pm}(\widehat x\cdot\omega) - 1),
\end{split}
\nonumber
\end{equation}
by Theorem 2.7 and (\ref{eq:Sect3Phivanish}) and Lemma 3.4, which of course holds with $\phi_{\pm}(x,\xi)$ replaced by $x\cdot\xi$, we have
$$
\int e^{-i\sqrt{\lambda}\omega\cdot x}T_{\pm}v_R^{(\pm)}dx \to 0.
$$
This combined with (\ref{eq:Sect4limitresoltoF}) proves the lemma. \qed


\begin{lemma}
Let $\rho(t)$ be as in Lemma 4.1, and $u_{\pm} = R(\lambda \pm i0)f, v_{\pm} = R(\lambda \pm i0)g$ with $f, g \in L^{2,s}$ for $s > 1/2$. Then
$$
\lim_{R\to\infty}\frac{\sqrt\lambda}{\pi R}\Big(\rho\big(\frac{|x|}{R}\big)
u_{\pm},v_{\pm}\Big) = \frac{1}{2\pi i}\left([R(\lambda + i0) - R(\lambda - i0)]f,g\right).
$$
\end{lemma}
Proof. By integration by parts
$$
\left((H-\lambda)\rho_1\big(\frac{r}{R}\big)u_{\pm},v_{\pm}\right) = \left(\rho_1\big(\frac{r}{R}\big)u_{\pm},g\right). 
$$
The left-hand side is equal to
$$
\left([H,\rho_1\big(\frac{r}{R}\big)]u_{\pm},v_{\pm}\right) + \left(\rho_1\big(\frac{r}{R}\big)f,v_{\pm}\right).
$$
Computing in the same way as in the previous lemma, we get the conclusion. \qed

\medskip
Recall the relation $\simeq$ defined by (\ref{eq:Sec2EquivRel}).


\begin{lemma}
For any $\phi \in L^2(S^1)$, we have
\begin{equation}
\int_{S^1}e^{i\sqrt{\lambda}\omega\cdot x}\chi_{\pm}(\widehat x \cdot\omega)\phi(\omega)d\omega
\simeq \left(\frac{2\pi}{\sqrt{\lambda}}\right)^{1/2}r^{-1/2}
e^{\pm i(\sqrt{\lambda}r - \frac{\pi}{4})}
\phi(\pm \widehat x).
\nonumber
\end{equation}
\end{lemma}
Proof. If $\phi \in C^{\infty}(S^1)$, this lemma follows from the stationary phase method. In the general case, we approximate $\phi$ by smooth function  and use $\mathcal F_0^{(\pm)}(\lambda)^{\ast} \in {\bf B}(L^2(S^1);\mathcal B^{\ast})$, which follows from Lemma 3.4. \qed


\begin{theorem}
For $\lambda > 0$ and $f \in \mathcal B$, the following asymptotic expansion holds:
$$
R(\lambda \pm i0)f \simeq \left(\frac{\pi}{\sqrt{\lambda}}\right)^{1/2}r^{-1/2}e^{\pm i(\sqrt{\lambda}r + \frac{\pi}{4})}\left(\mathcal F^{(\pm)}(\lambda)f\right)(\pm\omega),
$$
where $r = |x|, \omega = x/r$.
\end{theorem}
Proof. Since both side are bounded operators from $\mathcal B$ to $\mathcal B^{\ast}$, we have only to prove the theorem for $f \in C_0^{\infty}(\Omega)$. Let $\rho(t)$ be as in Lemma 4.1, and put $u_{\pm} = R(\lambda \pm i0)f$, $\phi_{\pm} = \mathcal F_{\pm}(\lambda)f$. Letting
$$
w_{\pm} = \pm \frac{i}{\sqrt2}\int_{S^1}e^{i\sqrt{\lambda}\omega\cdot x}
\chi_{\pm}(\widehat x\cdot\omega)\phi_{\pm}(\omega)d\omega
$$
and in view of Lemmas 2.1 and 4.3, we have only to prove
$u_{\pm} \simeq w_{\pm}$, namely
\begin{equation}
\frac{1}{R}\left(\rho\big(\frac{r}{R}\big)(u_{\pm} - w_{\pm}),u_{\pm} - w_{\pm}\right) \to 0.
\nonumber
\end{equation}
This is equivalent to showing that the following term tends to 0:
\begin{equation}
\begin{split}
& \frac{1}{R}\left(\rho\big(\frac{r}{R}\big)u_{\pm},u_{\pm}\right) + \frac{\pi}{\sqrt\lambda}\|\phi_{\pm}\|^2_{L^2(S^1)} \\
& \pm \frac{i}{\sqrt2R}\int_{S^1}\left(\int\rho\big(\frac{r}{R}\big)e^{-i\sqrt{\lambda}\omega\cdot x}\chi_{\pm}(\widehat x\cdot\omega)u_{\pm}dx\right)\overline{\phi_{\pm}(\omega)}\, d\omega + (CC),
\end{split}
\nonumber
\end{equation}
where $(CC)$ means the complex conjugate of the preceeding term.
By Lemmas 4.1, 4.2 and (\ref{eq:Parseval}), this converges to 0. \qed


\begin{theorem}
Let $\lambda > 0$ and $\phi \in L^2(S^1)$. Then
\begin{equation}
\left(\mathcal F^{(-)}(\lambda)^{\ast}\phi\right)(x) \simeq 
\frac{e^{\pi i/4}}{2\sqrt{\pi}\lambda^{1/4}}\left(
\frac{e^{-i\sqrt{\lambda}r}}{r^{1/2}}\phi(-\omega)
 - i \frac{e^{i\sqrt{\lambda}r}}{r^{1/2}}(\widehat S(\lambda)\phi\big)(\omega)\right),
\end{equation}
where $r = |x|, \omega = x/r$.
\end{theorem}
Proof. Since both side are bounded operators from $L^2(S^1)$ to $\mathcal B^{\ast}$, we have only to show the theorem for $\phi \in C^{\infty}(S^1)$. By (\ref{eq:Sect3MathcalFplusminus}), $\mathcal F^{(-)}(\lambda)^{\ast}\phi = \widetilde{\mathcal F_{0-}}(\lambda)^{\ast}\phi - R(\lambda + i0)\mathcal G_-(\lambda)^{\ast}\phi$. We apply the stationary phase method to the 1st term of the right-hand side, and Theorem 4.4 and Lemma 3.4 to the 2nd term. \qed


\subsection{Aharonov-Bohm solutions and the S-matrix}
The magnetic flux is defined by
\begin{equation}
\alpha = \frac{1}{2\pi}\lim_{r\to\infty}\int_{|x|=r}A.
\label{eq:flux}
\end{equation}

 Let $A^0 = \alpha(-x_2,x_1)/|x|^2$ and 
\begin{equation}
H_{AB} = (- i\nabla - A^0)^2 \quad {\rm in} \quad L^2({\bf R}^2).
\nonumber
\end{equation}
Its spectral properties are studied in \cite{Rui83} and \cite{ItoTa01}.
For $x \in {\bf R}^2$, let $\gamma(x;\omega)$ be the azimuth angle of $x$ to the direction $\omega \in S^1$ taking into account of the standard orientation.
Let $\psi_{AB}^{(\pm)}(x,\lambda,\omega)$ be the Aharonov-Bohm solution 
(see \cite{Rui83} and \cite{ItoTa01})
\begin{equation}
\psi_{AB}^{(\pm)}(x,\lambda,\omega) = \sum_{l\in{\bf Z}}\exp(\pm i|l - \alpha|\pi /2)\exp(il\gamma(x;\pm\omega))J_{|l-\alpha|}(\sqrt{\lambda}|x|).
\label{eqABsolution}
\end{equation}
It satisfies the Schr{\"o}dinger equation $\left(H_{AB} - \lambda\right)\psi_{AB}^{(\pm)} = 0$.
The Fourier transformation associated with $H_{AB}$ is defined by
\begin{equation}
\left(\mathcal F_{AB}^{(\pm)}f\right)(\lambda,\omega) = \frac{1}{2\sqrt2\pi}\int_{{\bf R}^2}
\overline{\psi_{AB}^{(\pm)}(x,\lambda,\omega)}f(x)dx,
\label{eq:FourierHAB}
\end{equation}
which is unitary $: L^2({\bf R}^2) \to L^2((0,\infty);L^2(S^1);d\lambda)$ and diagonalizes $H_{AB}$. 
It is worth recalling the following lemma proved by \cite{ItoTa01} illustrating the difference  of the spatial asymptotics between this distorted plane wave and the standard case, although we do not use it in this paper.


\begin{lemma} 
$\psi_{AB}^{(\pm)}$ is  bounded in ${\bf R}^2$, and has the asymptotic expansion
\begin{equation}
\psi_{AB}^{(\pm)}(x,\lambda,\omega) \sim \exp\big(i\alpha(\gamma(x;\mp\omega) - \pi)\big)e^{i\sqrt\lambda\omega\cdot x} 
+ \frac{e^{\mp i\sqrt{\lambda}r}}{r^{1/2}}\sum_{n=0}^{\infty}
\frac{c_n^{(\pm)}(\lambda,\widehat x,\omega)}{r^n}
\nonumber
\end{equation}
 in the region $|\widehat x \pm \omega| > \delta > 0$. Here $0 < \delta < 1$ is an arbitrarily fixed constant.
\end{lemma}

From here until the end of this section, we assume that for some $\epsilon_0 > 0$
\begin{equation}
|V(x)| \leq C\langle x\rangle^{-3/2-\epsilon_0}.
\label{eq:S4Vdecay3/2}
\end{equation}
We put
\begin{equation}
\Psi_{\pm}(x,\lambda,\omega) = \chi_{\Omega}(x)\psi_{AB}^{(\pm)}(x,\lambda,\omega) - R(\lambda \mp i0)\big((H - \lambda)\chi_{\Omega}(x)\psi_{AB}^{(\pm)}(x,\lambda,\omega)\big),
\label{eq:GenEigenfunction}
\end{equation}
$\chi_{\Omega}(x)$ being defined before Theorem 3.2. Since
$(H - \lambda)\chi_{\Omega}\psi_{AB}^{(\pm)} = O(r^{-3/2-\epsilon_0})$ by the asumption (\ref{eq:S4Vdecay3/2}), $\Psi_{\pm}$ is well-defined. 


\begin{theorem} (1) $\Psi_{\pm}$ satisfies $(H - \lambda)\Psi_{\pm} = 0$ in $\Omega$, $\Psi_{\pm} = 0$ on $\partial\Omega$. \\
\noindent
(2) 
For $f \in C_0^{\infty}(\Omega)$, we have
\begin{equation}
\big(\mathcal F^{(\pm)}(\lambda)f\big)(\omega)= \frac{1}{2\sqrt{2}\pi}\int_{\Omega}
\overline{\Psi_{\pm}(x,\lambda,\omega)}f(x)dx.
\label{eq:Fouriereigenfunction}
\end{equation}
\end{theorem}
Proof. The assertion (1) is obvious. 
The assertion (2) follows from general results of scattering theory and eigenfunction expansion theorem. In fact, letting $H_0 = - \Delta$ in $L^2({\bf R}^2)$, we define wave operators 
\begin{equation}
\begin{split}
W_{\pm}(H_{AB};H_{0}) & = \mathop{\rm s-lim}_{t\to\pm\infty}e^{itH_{AB}}e^{-itH_{0}}, \\
W_{\pm}(H;H_{0}) & = \mathop{\rm s-lim}_{t\to\pm\infty}e^{itH}r_{\Omega}e^{-itH_{0}}, \\
W_{\pm}(H;H_{AB}) & = \mathop{\rm s-lim}_{t\to\pm\infty}e^{itH}r_{\Omega}e^{-itH_{AB}}. 
\end{split}
\nonumber
\end{equation}
Then by the chain rule
\begin{equation}
W_{\pm}(H,H_{AB}) = W_{\pm}(H;H_0)W_{\pm}(H_0;H_{AB}).
\label{eq:chainrule}
\end{equation}
In \cite{Rui83} it is shown  that
$$
W_{\pm}(H_0;H_{AB})^{\ast} = W_{\pm}(H_{AB};H_{0}) = \left(\mathcal F_{AB}^{(\pm)}\right)^{\ast}\mathcal F_0.
$$
Therefore by Lemma 3.7 and (\ref{eq:chainrule}),
$W_{\pm}(H;H_{AB}) = \left(\mathcal F^{(\pm)}\right)^{\ast}\mathcal F_{AB}^{(\pm)}$, hence 
\begin{equation}
\mathcal F^{(\pm)} = \mathcal F_{AB}^{(\pm)}W_{\pm}(H_{AB};H).
\label{eq:FFABW}
\end{equation}
 This coincides with the Fourier transformation constructed by the perturbation method, which is just the formula (\ref{eq:Fouriereigenfunction}). \qed

\medskip
We define the scattering operators and their Fourier transforms by
\begin{equation}
\begin{split}
S(H;H_0) & = W_+(H;H_0)^{\ast}W_-(H;H_0), \\
\widehat S(H;H_0) &= \mathcal F_0S(H;H_0)\left(\mathcal F_0\right)^{\ast} = 
\mathcal F^{(+)}\left(\mathcal F^{(-)}\right)^{\ast},\\
S(H_{AB};H_0) &= W_+(H_{AB};H_0)^{\ast}W_-(H_{AB};H_0), \\
\widehat S(H_{AB};H_0)&= \mathcal F_0S(H_{AB};H_0)\left(\mathcal F_0\right)^{\ast} = 
\mathcal F_{AB}^{(+)}\left(\mathcal F_{AB}^{(-)}\right)^{\ast},\\
S(H;H_{AB}) &= W_+(H;H_{AB})^{\ast}W_-(H;H_{AB}), \\
\widehat S(H;H_{AB}) &= \mathcal F_{AB}^{(-)}S(H;H_{AB})\left(\mathcal F_{AB}^{(-)}\right)^{\ast}.
\end{split}
\nonumber
\end{equation}
By (\ref{eq:chainrule}) and (\ref{eq:FFABW}), one can prove the following lemma.


\begin{lemma}
$\ \widehat S(H;H_{AB}) = \widehat S(H_{AB};H_0)^{\ast}\widehat S(H;H_0)$.
\end{lemma}

$\widehat S(H;H_{AB})$ has the direct integral representation:
\begin{equation}
\widehat S(H;H_{AB}) = \int_0^{\infty}\oplus \, \widehat S(H,H_{AB};\lambda)d\lambda,
\nonumber
\end{equation}
where $\widehat S(H,H_{AB};\lambda)$ is a unitary operator on $L^2(S^1)$ called the S-matrix associated with $H$ and $H_{AB}$. Similarly, we define the S-matrices $\widehat S(H_{AB},H_0;\lambda)$ and $\widehat S(H,H_0;\lambda)$. Lemma 4.8 implies


\begin{lemma} For any $\lambda > 0$,
$\widehat S(H_{AB},H_0;\lambda)\widehat S(H,H_{AB};\lambda) = \widehat S(H,H_0;\lambda).$
\end{lemma}

Define the scattering amplitude $F_{AB}(\lambda)$ by
\begin{equation}
\widehat S(H,H_{AB};\lambda) = 1 - 2\pi i\,F_{AB}(\lambda).
\nonumber
\end{equation}
Using (\ref{eq:GenEigenfunction}) and Theorem 4.4, one can prove the following lemma.


\begin{lemma}
 $\Psi_{-}$ has the asymptotic expansion
\begin{equation}
\Psi_{-}(x,\lambda,\omega) - \psi_{AB}^{(-)}(x,\lambda,\omega) \simeq 
 \frac{e^{i\sqrt{\lambda}r}}{r^{1/2}}f_{+}(\lambda,\widehat x,\omega),
\nonumber
\end{equation}
where $f_+(\lambda,\theta,\omega) = C(\lambda)F_{AB}(\lambda,\theta,\omega)$,  $C(\lambda) = (2\pi)^{3/2}\lambda^{-1/4}e^{-\pi i/4}$ and $F_{AB}(\lambda,\theta,\omega)$ is the integral kernel of $F_{AB}(\lambda)$:
$$
F_{AB}(\lambda,\theta,\omega) = \frac{1}{2\sqrt2\pi}\mathcal F^{(+)}(\lambda)(H-\lambda)\chi_{\Omega}\psi_{AB}^{(-)}.
$$
\end{lemma}

By \cite{Rui83}, $\widehat S(H_{AB},H_0;\lambda)$ has the following integral kernel
\begin{equation}
\left(S_{\alpha}\phi\right)(\theta) = \int_{-\pi}^{\pi}s_{\alpha}(\theta - \theta')\phi(\theta')d\theta',
\nonumber
\end{equation}
\begin{equation}
s_{\alpha}(\theta) = \delta(\theta)\cos(\pi\alpha) + i \frac{\sin(\pi\alpha)}{\pi}\,{\rm p.v.}\frac{e^{i[[\alpha]]\theta}}{1 - e^{i\theta}},
\nonumber
\end{equation}
where $\alpha$ is the magnetic flux defined by (\ref{eq:flux}) and $[[\alpha]]$ is the least integer greater than or equal to $\alpha$. (Note that $\alpha$ in this paper is $- \alpha$ in \cite{Rui83}.) Let us note that $S_{\alpha}$ depends only on $\alpha$.
The spectrum of $S_{\alpha}$ consists of two eigenvalues $e^{\pm i\pi\alpha}$ with eigenvector $e^{im\theta}$, $\forall m \geq \alpha$, for $e^{i\pi\alpha}$ and eigenvector $e^{im\theta}$, $\forall m < \alpha$, for $e^{-i\pi\alpha}$. 

Let $\widehat S(\lambda;\theta,\omega)$ be the integral kernel of $\widehat S(H,H_0;\lambda)$. Lemmas 4.9 and 4.10 imply the following lemma.


\begin{lemma}
$$
\widehat S(\lambda;\theta,\omega) = s_{\alpha}(\theta-\omega) - 2\pi i
\int_0^{2\pi} s_{\alpha}(\theta-\theta')F_{AB}(\lambda,\theta',\omega)d\theta'.
$$
\end{lemma}


\section{Inverse problem}

We return to our original assumption (A-1), (A-2), (A-3) and study the inverse problem.


\subsection{Restriction of generalized eigenfunctions to a curve}
Take $R > 0$ so that $\mathcal O \subset \{|x| < R -1\}$.  
Let $D_{int} = \{|x| < R\}\cap\Omega$ and  $D_{ext} = \{|x| > R\}$. 
Let ${\mathcal F}^{(-)}(\lambda)$ be the generalized Fourier transform defined in \S 3 on $\Omega$. We put $C = \{|x| = R\}$ and
\begin{equation}
\langle f,g\rangle = \int_{C} f(x)\overline{g(x)}dl.
\nonumber
\end{equation}


\begin{lemma}
 If $f \in L^2(C)$ satisfies
\begin{equation}
\langle f,\mathcal F^{(-)}(\lambda)^{\ast}\phi\rangle = 0, \quad \forall \phi \in L^2(S^1),
\nonumber
\end{equation} 
then $f = 0$, provided $\lambda$ is not a Dirichlet eigenvalue for $H$ in $D_{int}$.
\end{lemma}
Proof. 
Let $R(z) = (H - z)^{-1}$. As is well-known, $R(\lambda - i0)$ can be extended to a bounded operator from $L^2(C)$ to $H^{3/2}_{loc}(\Omega)$, which is denoted by $L^2(C) \ni f \to R(\lambda - i0)\delta_Cf$. Let $R_{0}(z) = (H_{0} - z)^{-1}$, and $\chi_{\Omega}(x) \in C^{\infty}({\bf R}^2)$ be such that $\chi_{\Omega}(x) = 0$ if $|x| < R - 1/2$, $\chi_{\Omega}(x) = 1$ if $|x| > R$. Using the resolvent equation
\begin{equation}
R(\lambda - i0)\chi_{\Omega} = \chi_{\Omega}R_0(\lambda - i0)-  R(\lambda - i0)\big([H,\chi_{\Omega}] + \chi_{\Omega}(H - H_0)\big)R_0(\lambda - i0),
\nonumber
\end{equation}
and looking at the behavior at infinity of $R(\lambda - i0)\delta_Cf$, one can extend
$\mathcal F^{(-)}(\lambda)$  also on $L^2(C)$, which is denoted by $\mathcal F^{(-)}(\lambda)\delta_C$. By Theorem 4.4,
\begin{equation}
u := R(\lambda - i0)\delta_Cf \simeq C(\lambda)r^{-1/2}e^{-i\sqrt{\lambda}r}
\mathcal F^{(-)}(\lambda)\delta_Cf.
\nonumber
\end{equation}
The assumption of the lemma implies $\mathcal F^{(-)}(\lambda)\delta_Cf = 0$. Therefore
\begin{equation}
\lim_{R\to\infty}\frac{1}{R}\int_{|x|<R}|u(x)|^2dx = 0.
\label{eq:RellichDecay}
\end{equation}
Let us note that for any $\varphi \in C_0^{\infty}(\Omega)$
\begin{equation}
\begin{split}
((H - \lambda)u,\varphi) &= (u,(H-\lambda)\varphi) \\
 & = \langle f, R(\lambda + i0)(H - \lambda)\varphi\rangle \\
 & = \langle f, \varphi\rangle,
\end{split}
\nonumber
\end{equation}
where we have used the fact that $\varphi = R(\lambda + i0)(H - \lambda)\varphi$, since $\varphi$ is compactly supported, hence satisfies the radiation condition. We then have  $(H - \lambda)u = 0$  outside and inside $C$. Using (\ref{eq:RellichDecay}), we have $u = 0$ outside $C$ by Theorem 2.5. Since $u \in H^{3/2}_{loc}({\Omega})$, $u\big|_C = 0$. Since $\lambda$ is not a Dirichelt eigenvalue, $u = 0$ in $D_{int}$. Therefore $u = 0$ globally in $\Omega$, which implies $f = 0$. \qed


\subsection{Dirichlet-Neumann map}
If $\lambda$ is not a Dirichlet eigenvalue, the boundary value problem
\begin{equation}
\left\{
\begin{split}
& \big((-i \nabla - A)^2 + V- \lambda\big)u = 0 \quad {\rm in} \quad D_{int}, \\
& u = f \in H^{3/2}(C) \quad {\rm on} \quad C = \{|x| = R\} \\
& u = 0  \quad {\rm on} \quad \partial\Omega
\end{split}
\right.
\nonumber
\end{equation}
 has a unique solution $u$. Let
\begin{equation}
\Lambda(A,\lambda) : f \to \Big(\frac{\partial u}{\partial\nu} - i\nu\cdot A u\Big)\Big|_{C}
\nonumber
\end{equation}
be the Dirichlet-Neumann map (D-N map), $\nu$ being the outer unit normal to $C$. 


\subsection{Gauge equivalence}
In the following, we shall assume that

\medskip
\noindent
(A-4)  $\ \ {\displaystyle A(x) = \alpha\frac{(-x_2,x_1)}{|x|^2} + A'(x) \quad {\rm in} \quad \Omega}$, 

\medskip
\noindent
where $\alpha \in {\bf R}$ is the magnetic flux defined by (\ref{eq:flux}), and $A'(x)$ satisfies
\begin{equation}
|\partial_x^{\alpha}A'(x)| \leq C_{\alpha}\langle x\rangle^{-1 - |\alpha| - \epsilon_0}, \quad
\forall \alpha.
\label{eq:A'estimate}
\end{equation}

The conditions (A-1) and (A-2) follow from (A-4).
 We put
\begin{equation}
\Theta(x) = \frac{x\times d{\vec x}}{|x|^2}, \quad
d{\vec x} = (dx_1,dx_2).
\label{eq:DefAngleform}
\end{equation}
Take $R > 0$ large enough, and for $x \in {\bf R}^2\setminus\{0\}$, let $C(x)$ be a $C^{\infty}$-curve emanating from $(R,0)$ with end point $x$. 
Put
$$
\theta(x) = \int_{C(x)}\Theta(x).
$$
Then $e^{i\theta(x)} = (x_1 + ix_2)/|x|$. Let $\mathcal L_{\infty,0}$ be the set of real-valued functions $L(x) \in C^{\infty}(\overline{\Omega})$ such that for some $\epsilon_0 > 0$
\begin{equation}
|\partial_x^{\alpha}L(x)| \leq C_{\alpha}\langle x\rangle^{-\epsilon_0-|\alpha|}, \quad \forall \alpha.
\label{eq:Sec4Lxdefine}
\end{equation}

Recall that $\Omega = {\bf R}^2\setminus\mathcal O$, and we assume that $(0,0) \not\in \Omega$. We define
\begin{equation}
R_0 = \sup_{x \in \mathcal O} |x|.
\nonumber
\end{equation}


\begin{definition}
The gauge group ${\bf G}(\Omega)$ is a set of ${\bf C}$-valued functions $g(x) \in C^{\infty}(\overline{\Omega})$ satisfying $|g(x)| = 1$ on $\Omega$ and there exist $n \in {\bf Z}$ and $L \in \mathcal L_{\infty,0}$ such that
$g(x) = \exp\left(in\,\theta(x) + iL(x)\right)$ for $|x| > R_0$ .
\end{definition}

By the above definition, $n$ of $g(x)$ is computed as
\begin{equation}
n = \lim_{R\to\infty}-i\int_{|x|=R}\frac{dg}{g}.
\label{S5nofg}
\end{equation}

Two vector potentials $A^{(1)}$ and $A^{(2)}$ are said to be {\it gauge-equivalent} if there exists $g \in {\bf G}(\Omega)$ such that, being identified with 1-form,
$$
A^{(2)} = A^{(1)} - ig^{-1}dg.
$$
Or, equivalently, if there exist $n \in {\bf Z}$ and $g(x) \in C^{\infty}(\overline{\Omega})$ such that $|g(x)| = 1$ and $g(x) = e^{in\theta(x)}g_1(x)$, 
$|\partial_x^{\alpha}\left(g_1(x) - 1\right)| \leq C_{\alpha}\langle x\rangle^{-|\alpha|-\epsilon_0}$ such that 
\begin{equation}
A^{(2)} = A^{(1)} + n\Theta(x) - ig_1^{-1}dg_1.
\label{eq:Sec5Agaugeeq}
\end{equation}

Let $H(A,V)$ be the Schr{\"o}dinger operator with magnetic vector potential $A$ and electric scalar potential $V$ satisfying the assumptions (A-3) and (A-4). Two such operators $H(A^{(1)},V)$ and $H(A^{(2)},V)$ are said to be {\it gauge equivalent} if there exists $g \in {\bf G}(\Omega)$ such that
\begin{equation}
H(A^{(2)},V) = g\,H(A^{(1)},V)\,g^{-1}.
\label{eq:Sect5Gaugeequiv}
\end{equation}
We also say that $H^{(1)}$ and $H^{(2)}$ are gauge equivalent by $g \in {\bf G}(\Omega)$.
Let $e^{in\theta(D_x)}$ be the $\Psi$DO with symbol $e^{in\theta(\xi)}$:
$$
\Big(e^{in\theta(D_x)}f\Big)(x) = (2\pi)^{-1}\int_{{\bf R}^2}e^{ix\cdot\xi}e^{in\theta(\xi)}\widehat f(\xi)d\xi.
$$
The following lemma is well-known (cf. for example \cite{Nic00}, \cite{We02}, \cite{Ya06}).


\begin{lemma} Suppose $H^{(1)} = H(A^{(1)},V)$ and $H^{(2)} = H(A^{(2)},V)$ are gauge equivalent by $g \in {\bf G}(\Omega)$. Then we have
\begin{equation}
\widehat S(H^{(2)};H_0) = e^{in\theta(D_x)}
\widehat S(H^{(1)};H_0)e^{-in(\theta(D_x) + \pi)},
\label{eq:SmatrixGaugeeq}
\end{equation}
where $n$ is given by (\ref{S5nofg}), and $n = \alpha_2 - \alpha_1$, $\alpha_j$ being the magnetic flux of $H^{(j)}$.
\end{lemma}
Proof. By (\ref{eq:Sect5Gaugeequiv}), we have
\begin{equation}
 e^{in\theta(x)}g_1(x)W_{\pm}(H^{(1)};H_0) 
= 
{\mathop{\rm s-lim}_{t\to\pm\infty}}\,
e^{itH^{(2)}}r_{\Omega}e^{in\theta(x)}g_1(x)e^{-itH_0}.
\nonumber
\end{equation}
Let $\widehat f(\xi) \in C_0^{\infty}({\bf R}^2\setminus\{0\})$. Then by the stationary phase method we have as $t \to \pm \infty$
\begin{equation}
\begin{split}
e^{in\theta(x)}g_1(x)e^{-itH_0}f &\sim 
e^{in\theta(x)}e^{-itH_0}f \\
& \sim \frac{C}{t}e^{\frac{i|x|^2}{4t}}e^{in\theta(x)}\widehat f\big(\frac{x}{2t}\big) \\
& \sim (2\pi)^{-1}\int_{{\bf R}^2}e^{i(x\cdot\xi - t|\xi|^2)}
e^{in\theta(\pm\xi)}\widehat f(\xi)d\xi,
\end{split}
\nonumber
\end{equation}
where in the last step we have used that $\theta(\xi)$ is homogeneous of degree 0. Then we have
\begin{equation}
g\,W_{\pm}(H^{(1)};H_0) = W_{\pm}(H^{(2)};H_0)e^{in\theta(\pm D_x)}.
\nonumber
\end{equation}
Since $\theta(- \xi) = \theta(\xi) + \pi$, we obtain the lemma.
 \qed

\medskip
We study the converse of Lemma 5.3. We say that two S-matrices $\widehat S(H^{(i)},H_0)$, $i = 1,2$, are gauge equivalent if 
\begin{equation}
\widehat S(H^{(2)};H_0) = e^{in\theta(D_x)}
\widehat S(H^{(1)};H_0)e^{-in(\theta(D_x) + \pi)},
\nonumber
\end{equation}
 holds for some integer $n$. In this case, letting $H^{(3)} = H(A^{(3)},V)$ with $A^{(3)} = A^{(1)} + n\Theta(x) - ig_1^{-1}dg_1$, we have
\begin{equation}
\widehat S(H^{(2)};H_0) = \widehat S(H^{(3)};H_0).
\nonumber
\end{equation}
In the following we assume that $\mathcal O \subset \{|x| < R-1\}$.
The following two lemmas were proved by Nicoleau and Weder (\cite{Nic00}, Theorem 1.7, \cite{We02} Theorem 1.4).


\begin{lemma}
Suppose
$ \widehat S(H(A^{(1)},V^{(1)});H_0) = \widehat S(H(A^{(2)},V^{(2)});H_0)$. Let $\{x_0 + s\omega \, ; s \in {\bf R}\}$ $(\omega \in S^1)$ be a line which does not intersect $B_{R-1} = \{ x \in {\bf R}^2 \, ; |x| < R - 1\}$. Then 
\begin{equation}
\exp\Big(i\int_{-\infty}^{\infty}A^{(1)}(x_0 + s\omega)\cdot\omega ds\Big) = 
\exp\Big(i\int_{-\infty}^{\infty}A^{(2)}(x_0 + s\omega)\cdot\omega ds\Big),
\label{eq:IntegralA}
\end{equation}
\begin{equation}
\int_{-\infty}^{\infty}V^{(1)}(x_0 + s\omega) ds = 
\int_{-\infty}^{\infty}V^{(2)}(x_0 + s\omega) ds.
\label{eq:IntegralV}
\end{equation}
\end{lemma}


\begin{lemma}
Suppose (\ref{eq:IntegralA}) and (\ref{eq:IntegralV}) hold. Let $\alpha_j$ be the magnetic flux of $A^{(j)}$, and decompose $A^{(j)}$ as 
$A^{(j)}(x) = \alpha_j(-x_2,x_1)/|x|^2 + {A^{(j)}}'(x)$. Assume that 
\begin{equation}
|{A^{(1)}}'(x) - {A^{(2)}}'(x)| \leq C_N\langle x\rangle^{-N}, \quad \forall N > 0,
\label{eq:A'rapiddecay}
\end{equation}
\begin{equation}
|V^{(1)}(x) - V^{(2)}(x)| \leq C_N\langle x\rangle^{-N}, \quad \forall N > 0.
\label{eq:Vrapiddecay}
\end{equation}
Then $\alpha_2 - \alpha_1$ is an even integer and
there exists $L_1 \in \mathcal L_{\infty,0}$ such that ${A^{(2)}}' = {A^{(1)}}' + dL_1$ for $|x| > R-1$. Moreover
\begin{equation}
V^{(1)}(x) = V^{(2)}(x) \quad {\rm for} \quad |x| > R-1.
\label{eq:Sect5V1=V2}
\end{equation}
\end{lemma}

The assumptions (\ref{eq:A'rapiddecay}) and (\ref{eq:Vrapiddecay}) are used when we apply the support theorem of the Radon transform (\cite{He99}, p. 10, \cite{Na01}, p. 30).

By extending $L_1(x)$ to be a $C^{\infty}({\overline\Omega})$-function, we get the following corollary, since the gauge transformation $A^{(1)} \to A^{(1)} + dL_1$ does not affect the scattering operator.


\begin{cor}
Suppose $\alpha_1 = \alpha_2$ and 
$\widehat S(H(A^{(1)},V^{(1)});H_0) = \widehat S(H(A^{(2)},V^{(2)});H_0)$
holds. Then there exists $L_1 \in {\mathcal L}_{\infty,0}$ such that if we let 
$A^{(3)} = A^{(1)} + dL_1$, we have $A^{(2)} = A^{(3)}$ and $V^{(1)} = V^{(2)}$ for $|x| > R-1$, hence $H(A^{(2)},V^{(2)}) = H(A^{(3)},V^{(1)})$ for $|x| > R-1$,  and
$\widehat S(H(A^{(3)},V^{(1)});H_0) = \widehat S(H(A^{(2)},V^{(2)});H_0)$.
\end{cor}

We are now in a position to state our main theorem. We consider two Schr{\"o}dinger operators $H(A^{(i)},V^{(i)})$ defined in a domain $\Omega^{(i)}$, $i = 1,2$, satisfying the assumptions (A-3) and (A-4).


\begin{theorem}
Assume that $\widehat S(H(A^{(1)},V^{(1)}),H_0) = \widehat S(H(A^{(2)},V^{(2)}),H_0)$, and 
(\ref{eq:A'rapiddecay}) and  (\ref{eq:Vrapiddecay}) are satisfied. Assume also $\alpha_1 = \alpha_2$.
Then $\Omega^{(1)} = \Omega^{(2)}$, i.e. the obstacles are the same and $A^{(1)}$ and $A^{(2)}$ are gauge equivalent. Moreover, $V^{(1)} = V^{(2)}$ on $\Omega^{(1)} = \Omega^{(2)}$.
\end{theorem}
Proof. Let $H^{(i)} = H(A^{(i)},V^{(i)})$. By Corollary 5.6, one can assume that $A^{(1)} = A^{(2)}$ and $V^{(1)}(x) = V^{(2)}(x)$ for $|x| > R-1$.

Let $u_j = \mathcal F_j^{(-)}(\lambda)^{\ast}\phi$, where $\phi \in L^2(S^1)$ and $\mathcal F_j^{(-)}$ is the spectral representation for $H^{(j)}$. Let $u = \mathcal F_1^{(-)}(\lambda)^{\ast}\phi - \mathcal F_2^{(-)}(\lambda)^{\ast}\phi$. Since $H^{(1)} = H^{(2)}$ for $|x| > R-1$, we have
$(H^{(1)} - \lambda)u = 0$ for $|x| > R-1$. Furthermore, in view of 
Theorem 4.5, we have
$$
\frac{1}{R}\int_{|x|<R}|u(x)|^2dx \to 0, \quad {\rm as}\quad R \to \infty.
$$
Then by Theorem 2.3 and the unique continuation theorem, $u=0$ for $|x| > R-1$.

Let $D_{int} = \{ |x| < R\}$ and $C = \{|x| = R\}$.
Let $\Lambda^{(i)}(\lambda)$ be the D-N map for $H^{(i)}$ on $D_{int}$. Here we assume that $\lambda$ is not a Dirichelt eigenvalue for $H^{(i)}$, $i = 1,2$. 
Letting $\nu$ be the unit normal on $C$, we then have
$$
\frac{\partial}{\partial\nu}\mathcal F^{(-)}_1(\lambda)^{\ast}\phi - \frac{\partial}{\partial\nu}
\mathcal F^{(-)}_2(\lambda)^{\ast}\phi = \Lambda^{(1)}(\lambda)\mathcal F^{(-)}_1(\lambda)^{\ast}\phi - \Lambda^{(2)}(\lambda)\mathcal F^{(-)}_2(\lambda)^{\ast}\phi = 0. 
$$
By Lemma 5.1, the range of $\mathcal F_i^{(-)}(\lambda)$ is dense in $L^2(C)$, 
which implies
\begin{equation}
\Lambda^{(1)}(\lambda) = \Lambda^{(2)}(\lambda),
\label{eq:EllipticDNequal}
\end{equation}
for all $\lambda > 0$ except for a discrete set.
Let us now consider the hyperbolic initial-boundary value problem
\begin{equation}
\left\{
\begin{split}
&\partial_t^2u + (- i \nabla_x - A^{(j)})^2u + V^{(j)}u = 0, \quad {\rm in} \quad \Omega \times (0,\infty), \\
& u = \partial_tu = 0 \quad {\rm for} \quad t = 0, \\
& u = 0, \quad {\rm on} \quad \partial{\mathcal O}_j, 
\quad j = 1, \cdots, N,
\end{split}
\right.
\nonumber
\end{equation}
By (\ref{eq:EllipticDNequal}), the associated hyperbolic D-N maps $\Lambda_H^{(j)}$ also coincide on $C\times(0,\infty)$.

It is well-known that by virtue of the BC-method, one can determine the domain $\Omega$ and the operator $(-i\nabla_x - A(x))^2 +  V(x)$ from the hyperbolic D-N map.
Namely the following theorem holds (see e.g. \cite{Be97}, \cite{KaKuLa01} or \cite{Es06}, \cite{Es07}).

\begin{theorem}
If hyperbolic D-N maps coincide on $C\times(0,\infty)$, then
 $\Omega^{(1)} = \Omega^{(2)}$, $V^{(1)} = V^{(2)}$, and $A^{(1)}$ and $A^{(2)}$ are gauge equivalent with the gauge $g(x)$ in $|x| < R$, which is equal to 1 on $|x| = R$. 
 \end{theorem}
 
 Extending $g(x)$ to be 1 for $|x| > R$, we get that $A^{(1)}$ and $A^{(2)}$ are gauge equivalent in $\Omega$. \qed
 
 \medskip
 Note that in  Theorem 5.7, $A^{(2)} = A^{(1)} - ig^{-1}dg$ where $g = 1 + O(|x|^{-\epsilon_0})$ as $|x| \to \infty$.

\medskip
  In view of Theorem 5.7, we arrive at a natural conjecture : For non-integer flux case, $\alpha_1 = \alpha_2$ if $\widehat S(H^{(1)},H_0) = \widehat S(H^{(2)},H_0)$. If this is true, Theorem 5.7 is formulated as follows. For the sake of simplicity, we state the case without electric scalar potential : {\it For non-integer flux case, $A^{(1)}$ and $A^{(2)}$ are gauge equivalent if and only if $\widehat S(H^{(1)},H_0)$ and $\widehat S(H^{(2)},H_0)$ are gauge equivalent.}

Concerning  this conjecture, let us consider a simple case when the obstacles are known to be equal and convex.


\begin{theorem}
Suppose $H(A^{(i)},V^{(i)})$, $i = 1, 2$, are two operators in the same domain $\Omega$, where the obstacle $\mathcal O = {\bf R}^2\setminus\Omega$ is bounded and convex. Suppose 
\begin{equation}
A^{(i)}(x) = \alpha_i\frac{(-x_2,x_1)}{|x|^2} + {A^{(i)}}',
\label{S5A=frac+A'}
\end{equation}
where $\alpha_i \not\in {\bf Z}$, and the assumption (A-3) and (\ref{eq:A'rapiddecay}), (\ref{eq:Vrapiddecay}) are satisfied. 
If $S(H(A^{(1)},V^{(1)});H_0) = S(H(A^{(2)},V^{(2)});H_0)$, then $\alpha_1 = \alpha_2$.
\end{theorem}

Proof. By Lemma 5.5, we have 
$$
V^{(2)} = V^{(1)}, \quad {A^{(2)}}' = {A^{(1)}}' + dL_1, \quad \alpha_2 - \alpha_1 = 2m,
$$
with an integer $m$. In Lemma 5.5 we were considering on the set $\{|x|>R-1\}$, however, the proof works also outside a convex set. We put
\begin{equation}
H(A^{(3)},V^{(3)}) = e^{-i(2m\theta +L_1)}H^{(1)}e^{i(2m\theta + L_1)}.
\label{S5H2H1gauhe}
\end{equation}
Note that $A^{(3)} = A^{(2)}$, $V^{(3)} = V^{(2)}$, i.e. 
$$
H^{(3)} := H(A^{(3)},V^{(3)}) = H(A^{(2)},V^{(2)}) =: H^{(2)}.
$$
This implies that $S^{(3)} = S^{(2)}$, where $S^{(i)}$ is the scattering operator for $H^{(i)}$. It then follows from (\ref{S5H2H1gauhe}) and Lemma 5.3 that
$$
S^{(3)} = e^{-i2m\theta}S^{(1)}e^{i2m\theta}.
$$
Since $S^{(1)} = S^{(2)}= S^{(3)}$, we get
\begin{equation}
S^{(2)} = e^{-i2m\theta}S^{(2)}e^{i2m\theta}.
\label{S5S2equal}
\end{equation}
We shall have a contradiction assuming that $m \neq 0$.

Let $S^{(2)}(\theta,\theta)$ be the distribution kernel of the operator $S^{(2)}$. By Roux-Yafaev (\cite{RoYa02}, \cite{RoYa03}, and \cite{Ya06}, Theorem 4.3), we have
\begin{equation}
S^{(2)}(\theta,\theta') = s_{2}(\theta - \theta') + s'_{2}(\theta,\theta'),
\label{S5KernelS2}
\end{equation}
where
\begin{equation}
s_{2}(\theta) = \cos(\pi\alpha_2) \delta(\theta) + \frac{i\sin(\pi\alpha_2)}{\pi}{\rm p.v.}
\frac{e^{[[\alpha_2]]\theta}}{1 - e^{i\theta}},
\label{salpha}
\end{equation}
\begin{equation}
|s'_{2}(\theta,\theta')| \leq C|\theta-\theta'|^{-\delta}, \quad 0 \leq \delta < 1.
\label{S5salpah'}
\end{equation}
Here lets us note that in \cite{RoYa02}, \cite{RoYa03}, there is no obstacle. 
However, the presence of the obstacle needs only a little modification. In fact, Theorem 4.3 of \cite{Ya06} is based on its Theorem 3.3, whose technical background is the estimates of the resolvent multiplied by pseudo-differential operators (micro-local resolvent estimates). In the case of the exterior problem, these micro-local resolvent estimates are extended in the following way. Let $R(z) = (H - z)^{-1}$ be the resolvent for the exterior problem. We extend $A(x)$ and $V(x)$ smoothly to whole ${\bf R}^2$ and let $\widetilde H$ be the associated Hamiltonian, and $\widetilde R(z) = (\widetilde H - z)^{-1}$. We take $\chi(x) \in C^{\infty}({\bf R}^2)$ such that $\chi(x) = 0$ in a neighborhood of the obstacle and $\chi(x) = 1$ near infinity. We then have
\begin{equation}
R(z)\chi = \chi\widetilde R(x) - R(z)[H,\chi]\widetilde R(z).
\label{S5Rsolventeq}
\end{equation}
Since $[H,\chi]$ is compactly supported, one can then extend micro-local resolvent estimates to $R(z)$ by a simple perturbation argument.
The proof of (\ref{S5salpah'}) is then same as \cite{Ya06}, Theorem 4.3.

The equality (\ref{S5KernelS2}) means that
$$
\left(e^{i2m(\theta - \theta')}-1\right)S^{(2)}(\theta,\theta') = 0.
$$
Therefore, $S^{(2)}(\theta,\theta') = 0$ on the open set where $e^{i2m(\theta - \theta')}-1 \neq 0$, in particular, when $|\theta - \theta'| > 0$ and small. 

Denote by $\Pi_{\epsilon}$ the following domain :
$$
\Pi_{\epsilon} = \{(\theta,\theta')\, ; \, a < \theta < b, \ \epsilon < \theta - \theta' < 2\epsilon\},
$$
where $a, b$ are fixed, and $\epsilon$ is small. It follows from (\ref{S5salpah'}) that 
$$
\int_{\Pi_{\epsilon}}s'_{2}(\theta,\theta')d\theta d\theta' \to 0, \quad \epsilon \to 0.
$$
On the other hand, we have on $\Pi_{\epsilon}$
$$
{\rm Re}\,s_2(\theta - \theta') = - \frac{\sin(\alpha_2\pi)}{\pi(\theta - \theta')}
+ O(1).
$$
Therefore
$$
- {\rm Re}\int_{\Pi_{\epsilon}}s_2(\theta-\theta')d\theta d\theta' = 
(b-a)\frac{\sin(\alpha_2\pi)}{\pi}\log 2 + o(1).
$$
Hence we have
$$
\int_{\Pi_{\epsilon}}S^{(2)}(\theta,\theta')d\theta d\theta' \neq 0,
$$
i.e. $S^{(2)}(\theta,\theta')$ is not zero when $\epsilon < |\theta - \theta'| < 2\epsilon$, $\epsilon$ is small. \qed

\medskip
The converse of Therprem 5.9 is also true.


\begin{theorem}
Let $H(A^{(i)},V^{(i)})$, $i = 1,2 $, satisfy (A-1), (A-2), (A-3) on the same doamin $\Omega$. Assume $S^{(1)} = S^{(2)}$ and $V^{(1)} = V^{(2)}$. Assume also $A^{(1)}$, $A^{(2)}$ are gauge equivalent and the fluxes are not integers. Then $\alpha_1 = \alpha_2$.
\end{theorem}

Proof. We use the same notation as in the proof of Theorem 5.9. Since $H^{(1)}$ and $H^{(2)}$ are gauge equivalent, there exists a gauge $g(x) \in {\bf G}(\Omega)$ such that $H^{(3)} = gH^{(1)}g^{-1} = H^{(2)}$. Note that
$$
g(x) = e^{i(2m\theta + L)}, \quad {\rm for}\quad |x| > R.
$$
Since $H^{(3)} = H^{(2)}$, we have $S^{(3)} = S^{(2)}$. Since we assume that $S^{(1)} = S^{(2)}$, we have that
$$
S^{(2)}(\theta,\theta') = e^{i2m\theta}S^{(1)}e^{-i2m\theta},
$$
i.e. we are in the same situation as in Theorem 5.9. Therefore by the same argument as in Theorem 5.9 proves Theorem 5.10. 
\qed

\medskip
Note that Theorem 5.10 implies that having $S^{(1)} = S^{(2)}$ the condition $\alpha_1 = \alpha_2$ is necessary for $H^{(1)}$ and $H^{(2)}$ to be gauge equivalent.

As for the convexity assumption of the obstacle in Theorem 5.9, we make a conjecture that we can remove it by assuming the smallness of $V^{(i)}$ and ${A^{(i)}}'$. We shall discuss it elsewhere.

\begin{remark}
Theorems 5.7 and 5.9 deal with magnetic potentials satisfying conditions (A-4), (\ref{eq:A'estimate}). Following Yafaev \cite{Ya06} (see also, \cite{BaWe07}, \cite{We02}, \cite{RoYa02}, \cite{RoYa03}), one can consider the class of magnetic potentials having the form $A(x) = A_0(x) + A'(x)$ for $|x| > R$, where $A_0(x) \in C^{\infty}({\bf R}^d\setminus\{0\})$, $d \geq 2$, and $A_0(x)$ is homogeneous of degree $- 1$, $A'(x)$ satisfies (\ref{eq:A'estimate}). It is assumed that $A_0(x)$ satisfies the transversality condition $x\cdot A_0(x) = 0$. In this case the gauge group $G(\overline{\Omega})$ consists of $g(x) \in C^{\infty}(\overline{\Omega})$, $|g(x)| = 1$ and $g(x) = e^{in\theta + i\varphi(\theta) + iL_1(x)}$ for $|x| > R$ in the case $d = 2$, where $n \in {\bf Z}$, $\varphi(\theta) \in C^{\infty}(S^1)$ and $L_1(x)$ satisfies (\ref{eq:Sec4Lxdefine}) (cf. Definition 5.2). In the case $d \geq 3$, $g(x) = e^{i\varphi(\theta) + iL_1(x)}$, where $\varphi(\theta) \in C^{\infty}(S^{d-1})$ and $L_1(x)$ satsifies (\ref{eq:Sec4Lxdefine}). Since the results of \S 2 $\sim$ \S 4 hold for this calss of potentials, one can show that analogues of Theorems 5.7 and 5.9 hold.
\end{remark}


\begin{thebibliography}{A}

\bibitem{AB59}
Y. Aharonov and D. Bohm, \textit{Significance of electromagnetic potential in the quantum theory}, Phys. Rev. \textbf{115} (1959), 485-491.

\bibitem{AgHo76} S. Agmon and L. H{\"o}rmander, \textit{Asymptotic
properties of solutions of differential equations with simple
characteristics},  J. d'Anal. Math. \textbf{30} (1976), 1-30.


\bibitem{BaWe07}
M. Ballesteros and R. Weder, \textit{High-velocity estimates for the scattering operator and the Aharonov-Bohm effect in three dimensions}, Commun. in Math. Phys. \textbf{285} (2009), 345-398.

\bibitem{Be97}
M. I. Belishev, \textit{Boundary control in reconstruction of manifolds and metrics (the BC method)}, Inverse Problems \textbf{13} (1997), R1-R45.

\bibitem{BeKu92}
I.M. Belishev and Y. Kurylev, \textit{The reconstruction of the Riemannian manifold via its spectral data}, Comm. in P. D. E. \textbf{17} (1992), 767-804.





\bibitem{Es03}
G. Eskin, \textit{Inverse problems for the Schr{\"o}dinger operators with electromagnetic potentials in domains with obstacles}, Inverse Problems \textbf{19} (2003), 985-996.

\bibitem{Es06}
G. Eskin, \textit{A new approach to hyperbolic inverse problems}, Inverse Problems \textbf{22} (2006), 815-831.


\bibitem{Es07}
G. Eskin, \textit{A new approach to hyperbolic inverse problems II (Global step)}, Inverse Problems \textbf{23} (2007), 2343-2356.



\bibitem{He99}
S. Helgason, \textit{The Radon transform}, 2nd edition, Birkh{\"a}user, Boston-Basel-Berlin (1999).

\bibitem{Ho85}
L. H{\"o}rmander, \textit{The Analysis of Linear Partial Differential Operators IV}, Springer-Verlag, Berlin Heidelberg New York Tokyo (1985).

\bibitem{IkSa72}
T. Ikebe and Y. Saito, \textit{Limiting absorption method and absolute continuity for the Schr{\"o}dinger operator}, J. of Math. Kyoto Univ. \textbf{12} (1972), 513-542.

\bibitem{IsKi85a}
H. Isozaki and H. Kitada, \textit{Modified wave operators with time-dependent modifiers}, J. Fac. Sci. Univ. Tokyo \textbf{32} (1985), 77-104.

\bibitem{IsKi85b}
H. Isozaki and H. Kitada, \textit{A remark on the micro-local resolvent estimates for two-body Schr{\"o}dinger operators}, Publ. RIMS Kyoto Univ. \textbf{21} (1985), 889-910.

\bibitem{Is01}
H. Isozaki, \textit{Asymptotic properties of solutions to 3-particle Schr{\"o}dinger equations }, Commun. in Math. Phys. \textbf{222} (2001), 371-413.

\bibitem{ItoTa01}
H. T. Ito and H. Tamura, \textit{Aharonov-Bohm effect in scattering by point-like magnetic fields at large separation}, Ann. Henri Poincar{\'e}  \textbf{2} (2001), 309-359.


\bibitem{JePe85}
A. Jensen and P. Perry, \textit{Commutator methods and Besov space estimates for Schr{\"o}dinger operators}, J. Operator Theory, \textbf{14} (1985), 181-188.

\bibitem{KaKuLa01}
A. Katchalov, Y. Kurylev and M. Lassas, \textit{Inverse Boundary Spectral Problems}, Chapman and Hall/CRC \textbf{123} (2001).

\bibitem{KuLa00} 
Y. Kurylev and M. Lassas, \textit{Hyperbolic inverse problems with data on a part of the boundary}, AMS/IP, Stud. Adv. Math. \textbf{16} (2000), 259-272.

\bibitem{LoTh87}
M. Loss and B. Thaller, \textit{Scattering of particles by long-range magnetic fields}, Ann. Phys. \textbf{176} (1987), 159-180.

\bibitem{Na01}
F. Natterer, \textit{The Mathematics of Computerized Tomography},  SIAM, Philadelphia (2001).

\bibitem{Nic00}
F. Nicoleau, \textit{An inverse scattering problem with the Aharonov-Bohm efdfect}, J. Math. Phys. \textbf{41} (2000), 5223-5237.

\bibitem{PeTo89}
M. Peshkin and A. Tonomura, \textit{The Aharanov-Bohm Effect}, Lect. Notes in Phys. \textbf{340}, Springer-Verlag, Berlin (1989).

\bibitem{RoYa02}
Ph Roux and D. Yafaev, \textit{On the mathematical theory of the Aharonov-Bohm effect}, J. Phys. A: Math. Gen. \textbf{35} (2002), 7481-7492.


\bibitem{RoYa03}
Ph Roux and D. Yafaev, \textit{The scattering matrix for the Schr{\"o}dinger operator with a long-range electro-magnetic potential}, J. Math. Phys. \textbf{44} (2003), 2762-2786.

\bibitem{Rui83}
S.N.M. Ruijsenaars, \textit{The Aharonov-Bohm effect and scattering theory}, 
Annals pf Phys. \textbf{146} (1983), 1-34.

\bibitem{Tam06}
H. Tamura, \textit{Semiclassical analysis for magnetic scattering by two solenoidal fields}, J. London Math. Soc. \textbf{74} (2006), 695-716.

\bibitem{Tam07}
H. Tamura, \textit{Semiclassical analysis for magnetic scattering by two solenoidal fields: total cross sections}, Ann. Henri Poincar{\'e} \textbf{8} (2007), 1071-1114.

\bibitem{Tetal86}
A. Tonomura, N. Osakabe, T. Matsuda, T. Kawasaki, J. Endo, S. Yano, and H. Yamada, \textit{Evidence for Aharonov-Bohm effect with magnetic field completely shielded from electron wave}, Phys. Rev. Lett. \textbf{56} (1986), 792.

\bibitem{We02}
R. Weder, \textit{The Aharonov-Bohm effect and time-dependent inverse scattering theory}, Inverse Problems \textbf{18} (2002), 1041-1056.

\bibitem{Ya03}
D. Yafaev, \textit{High-energy and smoothness asymptotic expansion of the scattering amplitude}, J. Funct. Anal. \textbf{202} (2003), 526-570.


\bibitem{Ya06}
D. Yafaev, \textit{Scattering by magnetic fields}, St. Petersburg Math. J.  \textbf{17} (2006), 675-695, Arxiv:math/0501544v.1.

\end{thebibliography}
\end{document}